\documentclass{emulateapj}

\input epsf.sty

\newcommand{\Msun}{M_{\odot}}

\def\gsim{\mathrel{\rlap{\lower 4pt \hbox{\hskip 1pt $\sim$}}\raise 1pt
\hbox {$>$}}}
\def\lsim{\mathrel{\rlap{\lower 4pt \hbox{\hskip 1pt $\sim$}}\raise 1pt
\hbox {$<$}}}

\begin{document}

\title{The Unique Type Ib Supernova 
2005bf at Nebular Phases: \\ 
A Possible Birth Event of A Strongly Magnetized Neutron Star
\altaffilmark{1}}

\author{
K.~Maeda\altaffilmark{2,3}, 
M.~Tanaka\altaffilmark{4}, 
K.~Nomoto\altaffilmark{4,5}, 
N.~Tominaga\altaffilmark{4}, 
K.~Kawabata\altaffilmark{6}, \\ 
P.A.~Mazzali\altaffilmark{2,4,7}, 
H.~Umeda\altaffilmark{4},  
T.~Suzuki\altaffilmark{4}, 
T.~Hattori\altaffilmark{8}}

\altaffiltext{1}{
Based on data collected at Subaru Telescope, which is operated by the 
National Astronomical Observatory of Japan.}
\altaffiltext{2}{Max-Planck-Institut f\"ur Astrophysik, 
Karl-Schwarzschild-Stra{\ss}e 1, 85741 Garching, Germany: 
maeda@MPA-Garching.MPG.DE}
\altaffiltext{3}{Department of Earth Science and Astronomy,
Graduate School of Arts and Science, University of Tokyo, 
3 - 8 - 1 Komaba, Meguro-ku, Tokyo
153-8902, Japan}
\altaffiltext{4}{Department of Astronomy, School of Science,
University of Tokyo, Bunkyo-ku, Tokyo 113-0033, Japan}
\altaffiltext{5}{Research Center for the Early Universe, School of
Science, University of Tokyo, Bunkyo-ku, Tokyo 113-0033, Japan}
\altaffiltext{6}{Hiroshima Astrophysical Science Center, Hiroshima University, Hiroshima, Japan}
\altaffiltext{7}{Instituto Nazionale di Astrofisica 
(INAF)-Osservatorio Astronomico di Trieste, Via Tiepolo 11, 
I-34131 Trieste, Italy}
\altaffiltext{8}{Subaru Telescope, National Astronomical Observatory 
of Japan, Hilo, HI 96720, USA} 

\begin{abstract}
Late phase nebular spectra and photometry 
of Type Ib Supernova (SN) 2005bf 
taken by the Subaru telescope 
at $\sim 270$ and $\sim 310$ days since the explosion are presented. 
Emission lines ([OI] $\lambda\lambda$6300, 6363, [CaII] 
$\lambda\lambda$7291, 7324, [FeII] $\lambda$7155) show 
the blueshift of $\sim 1,500 - 2,000$ km s$^{-1}$. 
The [OI] doublet shows a doubly-peaked profile. 
The line luminosities can be 
interpreted as coming from a blob or jet 
containing only $\sim 0.1 - 0.4\Msun$, in which $\sim 0.02 - 0.06\Msun$ 
is $^{56}$Ni synthesized at the explosion. 
To explain the blueshift, the blob should either be of unipolar 
moving at the center-of-mass velocity $v \sim 2,000 - 5,000$ km s$^{-1}$, 
or suffer from self-absorption within the ejecta as seen in SN 1990I. 
In both interpretations, 
the low-mass blob component dominates the optical output both at the first peak 
($\sim 20$ days) and at the late phase ($\sim 300$ days).  
The low luminosity at the late phase (the absolute $R$ magnitude $M_{R} 
\sim -10.2$ mag at $\sim 270$ days) 
sets the upper limit for the mass of $^{56}$Ni $\lsim 0.08\Msun$, 
which is in contradiction to the value necessary to explain the 
second, main peak luminosity ($M_{R} \sim -18.3$ mag 
at $\sim 40$ days). 
Encountered by this difficulty in the $^{56}$Ni heating model, 
we suggest an alternative scenario in which the heating source is 
a newly born, strongly magnetized neutron star (a magnetar) with the surface magnetic field 
$B_{\rm mag} \sim 10^{14-15}$ gauss and the initial spin period $P_{0} \sim 10$ ms. 
Then, SN 2005bf could be a link between normal SNe Ib/c and an X-Ray Flash associated SN 2006aj, 
connected in terms of $B_{\rm mag}$ and/or $P_0$. 
\end{abstract}


\section{INTRODUCTION}

SN 2005bf has been claimed to be extremely peculiar from 
the very beginning. 
The following features of SN 2005bf fall short of any expectations 
obtained from observations of past Type Ib/c supernovae (SNe Ib/c). 
(1) Discovered on 2005 April 6 (UT) 
by Monard (2005) and Moore \& Li (2005), 
it first showed no strong He lines although there 
was evidence of H$_{\alpha}$. Thereafter 
He lines were increasingly developed with time, so 
it then was classified as Type Ib 
(Anupama et al. 2005; Wang \& Baade 2005; 
Modjaz, Kirshner, \& Challis 2005). 
(2) The He lines show peculiar temporal evolution: 
the velocity increased with time (Tominaga et al. 2005).  
(3) The optical light curve is very unique 
showing double-peaks at $t \sim 20^{\rm d}$ and $\sim 40^{\rm d}$. 
It was brighter at the second peak, 
reaching the absolute bolometric magnitude 
$M_{\rm bol} \sim -18$ mag. 
Hereafter $t$ is the age of the supernova 
since the putative explosion date, which is taken as 
2005 March 28 (Tominaga et al. 2005). 
(4) Even more peculiarly, it 
declines very quickly after the second peak, 
nearly 2 magnitudes just in the subsequent 40 days. 
This rapidly fading light curve has never been 
observed in supernovae possibly 
except for another very peculiar SN Ic 1999as (Hatano et al. 2001). 
(5) The peak magnitude $M_{\rm bol} \sim -18$ mag is quite 
bright for the relatively late peak date at $t \sim 40^{\rm d}$. 
It requires $M$($^{56}$Ni) $\gsim 0.3\Msun$ 
in the usual $^{56}$Ni heating scenario for SNe Ib/c. 
Hereafter $M$($^{56}$Ni) is the mass of 
$^{56}$Ni synthesized at the explosion.  
For the summary of the early phase observations, 
see Anupama et al. (2005), Tominaga et al. (2005), and 
Folatelli et al. (2006). 

\begin{deluxetable*}{llll}[tb]
 \tabletypesize{\scriptsize}
 \tablecaption{Notation\tablenotemark{a} and Model Values\tablenotemark{b}
 \label{tab:model_01}}
 \tablewidth{0pt}
 \tablehead{
   \colhead{Epoch}
 & \colhead{$M_{\rm ej}$}
 & \colhead{$E_{51} \equiv E/10^{51}$ erg}
 & \colhead{$M$($^{56}$Ni)}
}
\startdata
$\lsim 80^{\rm d}$  & $M_{\rm ej, peak} \sim 7.3\Msun$        &  $E_{{\rm peak}, 51} \sim 1.3$ & 
$M$($^{56}$Ni)$_{\rm peak} \sim 0.32\Msun$\\
$\sim 300^{\rm d}$ & $M_{\rm ej, neb} \sim 0.12 - 0.34\Msun$ &  $E_{{\rm neb}, 51} 
\sim 0.015 - 0.085$ & 
$M$($^{56}$Ni)$_{\rm peak} \sim 0.024 - 0.056\Msun$\\
\enddata
\tablenotetext{a}{The subscript "peak" is used for the values derived by modeling the observations 
at $t \lsim 80^{\rm d}$ (Tominaga et al. 2005), and the subscript "neb" is for those derived 
by the nebular observations at $t \sim 300^{\rm d}$ (this work). }
\tablenotetext{b}{$M_{\rm ej}$ and $E$ are the mass and kinetic energy of the ejecta. 
$M$($^{56}$Ni) is the mass of $^{56}$Ni synthesized at the explosion.}
\end{deluxetable*}

Tominaga et al. (2005) tried to constrain the explosion physics 
and the progenitor of SN 2005bf by modeling the light curve and 
the spectra up to $t \sim 80^{\rm d}$. 
They used the distance modulus $\mu = 34.5$ and $E(B - V) = 0.045$, 
which we also adopt in this paper. 
In their best model, the supernova has massive ejecta 
($M_{\rm ej, peak} \sim 6 - 7\Msun$), 
normal kinetic energy 
($E_{\rm peak, {51}} \equiv E_{\rm peak}/10^{51}$ erg $\sim 1 - 1.5$), 
and relatively large $M$($^{56}$Ni)$_{\rm peak}$ $\sim 0.3\Msun$. 
In this paper, the subscript "peak" is used for 
the values derived by modeling the early phase observations 
(see Table 1). 
The model yields a good agreement with the observations 
{\it if} gamma-rays can escape more easily than in 
usual situation (i.e., {\it if} the opacity for gamma-rays, 
$\kappa_{\gamma}$, is decreased by a factor of 
$\sim 30$ from the canonical value). 
These values suggest SN 2005bf is from 
a WN star with the zero-age 
main-sequence mass $M_{\rm ms} \sim 25 - 30\Msun$. 
A similar conclusion was obtained independently 
by Folatelli et al. (2006), who also assumed artificially 
small $\kappa_{\gamma}$. 

\begin{figure*}[tb]
\begin{center}
	\begin{minipage}[t]{0.6\textwidth}
		\epsscale{1.0}
		\plotone{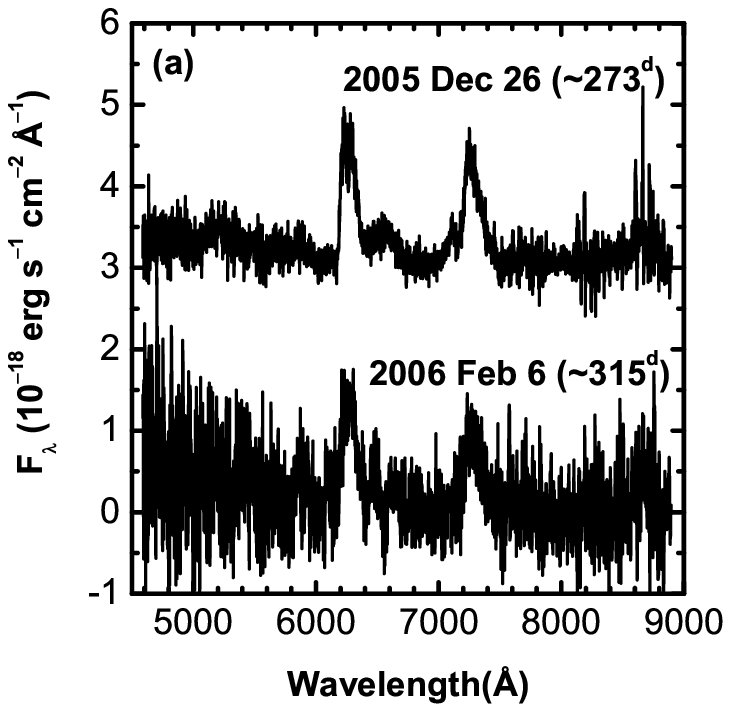}
	\end{minipage}
	\begin{minipage}[t]{0.4\textwidth}
		\epsscale{1.0}
		\plotone{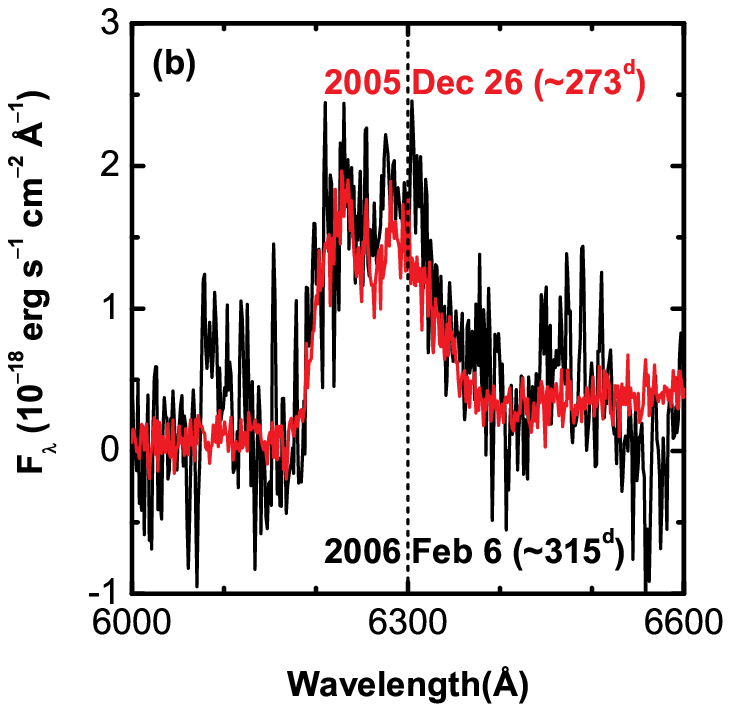}
	\end{minipage}
	\begin{minipage}[t]{0.4\textwidth}
		\epsscale{1.0}
		\plotone{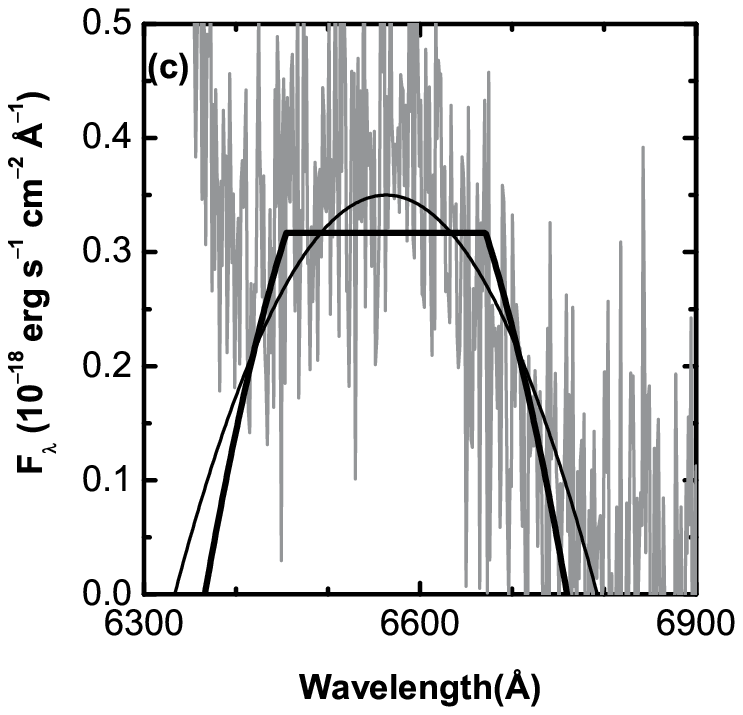}
	\end{minipage}
\end{center}
\caption[]
{The reduced spectra of SN 2005bf. The redshift of the 
host galaxy ($z = 0.018913$) is corrected for. 
The flux of the December spectrum is calibrated 
using the $R$-band photometry. 
(a) The spectra at 2005 December 26 (upper) and at 2006 Feb 6 (lower). 
The flux of the December spectrum is shifted upward 
by the amount of $3 \times 10^{-18}$ erg s$^{-1}$ cm$^{-2}$ 
\AA$^{-1}$ for presentation. (b) The expanded view at 6000 -- 6600\AA\ 
(the [OI] $\lambda\lambda$ 6300, 6363) 
for the December (red) and the February (black) spectra. The flux of the 
February spectrum is multiplied by an arbitrary amount for presentation. 
(c) The expanded view at 6300 -- 6900\AA\ of the December spectrum (gray). 
The parabola fit as the H$_{\alpha}$ emission is shown 
for the outermost velocity 10,500 km s$^{-1}$ (FWHM $\sim 15,000$ km s$^{-1}$; 
thin black). 
Also shown is the fit by the parabola with the outermost velocity 
9,000 km s$^{-1}$ with the central flat part below 5,000 km s$^{-1}$ 
(thick black). 
\label{fig1}}
\end{figure*}

In this paper, we present results from late-phase 
spectroscopy and photometry of SN 2005bf at 
$t \sim 273^{\rm d}$ and $315^{\rm d}$. 
SN 2005bf has clearly entered into the nebular phase, 
so it is possible to derive information qualitatively different from 
that derived with the early phase observations. 
The observed features turn out to be 
even more peculiar than expected from the early phase 
observations. We will critically examine some ideas 
whether they give a view consistently explaining both the 
previous and the new observations. 
At the end, we suggest a scenario that 
SN 2005bf is a birth event of a strongly magnetized neutron star 
(magnetar) and this central remnant is the heating source  
-- a scenario which could solve the puzzles found by our new observations. 

In \S 2, we describe the observation and data reduction. 
In \S 3, the nebular spectra are examined in detail. 
In \S 4, we present the light curve connecting the new and previous observations, 
and discuss a problem brought by the new observations. 
In \S 5, we discuss and critically examine possible underlying scenarios. 
Among the scenarios, we highlight the magnetar scenario in \S 6, 
where consequences and implications of this scenario are mentioned.
The paper is closed in \S 7 with conclusions.

\section{OBSERVATIONS AND DATA REDUCTION}

Spectroscopy and photometry of SN 2005bf have been
performed on 2005 December 26 (UT) and on 2006 February 6
with the 8.2 m Subaru telescope equipped with the
Faint Object Camera and Spectrograph
(FOCAS; Kashikawa et al. 2002). 
The epochs correspond $t \sim 273^{\rm d}$ and $\sim 315^{\rm d}$. 
For spectroscopy, we used $0\farcs 8$ width slit
and B300 grism, which gave a wavelength coverage
of 4700--9000 \AA\ and a spectral resolution of
$\simeq 10.7$ \AA . The exposure times were
12600 s and 6600 s for 2005 December and 2006
February, respectively. BD +28$^{\circ}$4211 and
G191B2B (Massey et al. 1988; Massey \& Gronwall 1990)
were also observed for flux calibrations.
For photometry, we obtained 180 s exposure images
with either $B$- or $R$-band filter on both nights.
The derived magnitudes were $B>25.6$ mag and $R=24.4\pm 0.2$ 
mag on 2005 December 26 and $B>24.6$ mag and $R>24.5$ mag on 2006
February 6. Since we could not recognize SN~2005bf in the
$B$-band image on 2005 December 26 and in the $B$- and $R$-band
images on 2006 February 6, we adopted $5-\sigma$ background
as the upper-limit of the magnitude.
We obtained images of standard stars around PG 0942-029
(Landolt 1992) for photometric calibrations.

Figure 1 shows the reduced spectra of SN 2005bf. 
At 2005 December 26, 
SN 2005bf was already in a nebular phase,  
characterized by strong emission lines with almost no continuum. 
No significant evolution is seen between 
December 26 and February 6 either in line profiles or line flux 
ratios, although the low S/N in the February spectrum prevents us 
from rejecting possible difference in detailed line structures. 
Spectroscopic features are discussed in \S 3 in detail.

Figure 2 shows the late phase $B$ (only upper limits) and $R$ magnitudes  
of SN 2005bf as combined with previously published ones 
(from Tominaga et al. 2005). 
The light curve is compared with the $R$-band and 
the bolometric light curves of SN Ic 1998bw (Patat et al. 2001), 
and with the $R$-band light curve of SN Ib 1990I 
(Elmhamdi et al. 2004) corrected for the distance and the reddening 
to the position of SN 2005bf. 
Surprisingly enough, SN 2005bf turned out to be extremely 
faint at the late epochs. 
The light curve characteristic is further discussed in \S 4, 
where we see that the faintness of SN 2005bf 
at the late epochs is difficult to understand 
in the context of a conventional supernova emission model.

\begin{figure*}[tb]
\begin{center}
	\begin{minipage}[t]{0.45\textwidth}
		\epsscale{1.0}
		\plotone{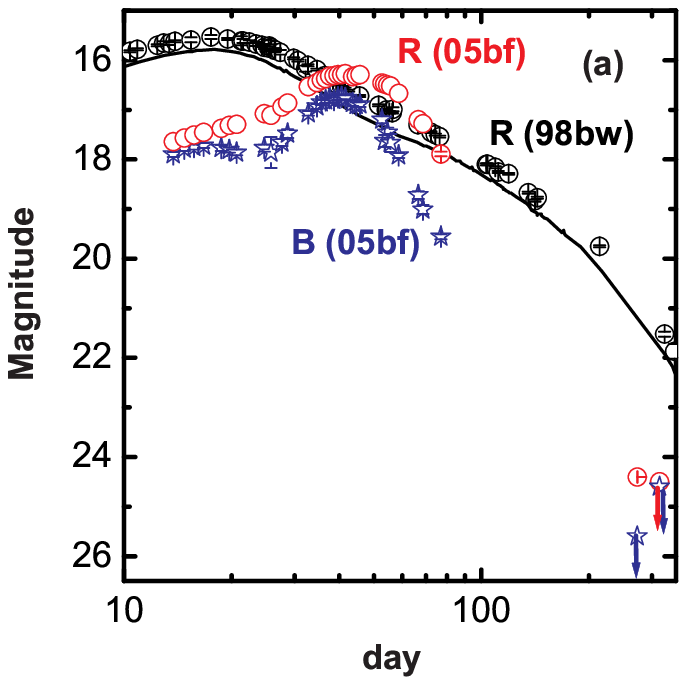}
	\end{minipage}
	\begin{minipage}[t]{0.45\textwidth}
		\epsscale{1.0}
		\plotone{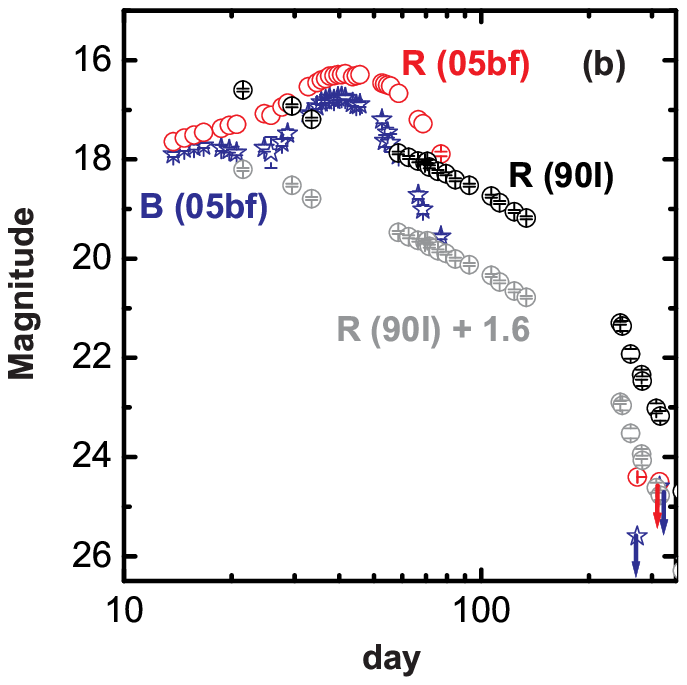}
	\end{minipage}
\end{center}
\caption[]
{The $R$-band (red open circles) and $B$-band (blue stars) light curves 
of SN 2005bf. 
Time is measured since the explosion, which is assumed to have occurred 
on 2005 March 28 (Tominaga et al. 2005). 
The Subaru observation of $R$ at 
December 26 and upper limits ($B$ at December 26, $R$ and $B$ at February 6) 
are shown. The early phase data ($< 100$ days) are from Tominaga et al. (2005). 
The light curves are compared with those 
of SN 1998bw (a: 
$R$ shown by black open circles, 
and the bolometric magnitude shown by a black curve; from Patat et al. 2001) and 
of SN 1990I (b: $R$ shown by black open circles; from Elmhamdi et a. 2004). 
The magnitudes of SNe 1998bw and 1990I are corrected for 
the distance modulus and the reddening to the position of SN 2005bf. 
\label{fig2}}
\end{figure*}

\begin{figure*}[tb]
\begin{center}
	\begin{minipage}[t]{0.45\textwidth}
		\epsscale{1.0}
		\plotone{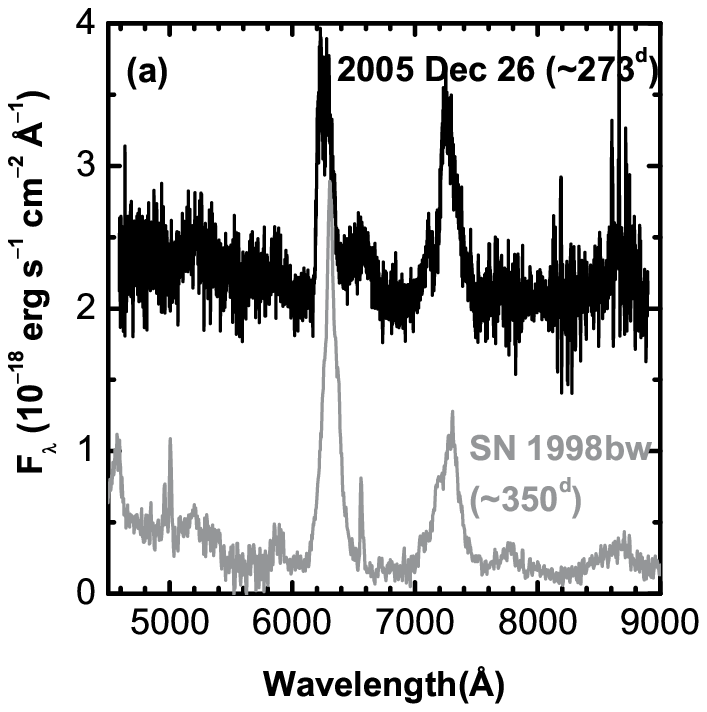}
	\end{minipage}
	\begin{minipage}[t]{0.45\textwidth}
		\epsscale{1.0}
		\plotone{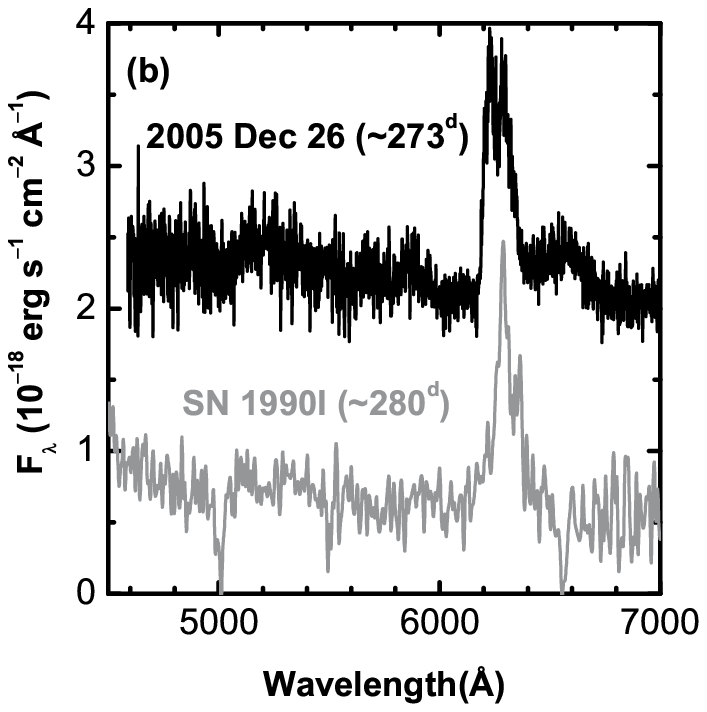}
	\end{minipage}
\end{center}
\caption[]
{Comparison of the December spectrum 
with (a) SN 1998bw and (b) SN 1990I at similar epochs. 
Flux is shifted artificially for SNe 1998bw and 1990I 
for presentation. 
The flux of the December spectrum is shifted upward 
by the amount of $2 \times 10^{-18}$ erg s$^{-1}$ cm$^{-2}$ 
\AA$^{-1}$ for presentation. 
The redshifts of the host galaxies are corrected for. 
\label{fig3}}
\end{figure*}

\section{NEBULAR SPECTRA}

\subsection{General Features}

The reduced spectra show strong emissions at 
$\sim 6300$\AA, $7300$\AA, which 
we interpret as [OI] $\lambda\lambda$6300, 6363 
doublet and [CaII] $\lambda$7300 
(actually a combination of [CaII] $\lambda\lambda$7291, 7324, 
and [FeII] $\lambda\lambda$7155, 7172, 7388, and 7452). 
Other emission features are marginally detected 
at $\sim 5200$\AA\ (likely a blend of [FeII]) 
and $\sim 8700$\AA (CaII IR and [CI] $\lambda$8727). 

A feature at $\sim 6,500$\AA\ is consistent with broad H$_{\alpha}$ emission 
(FWHM $\sim 15,000$ km s$^{-1}$ measured in the December spectrum). 
This feature was reported in a spectrum taken at 
2005 October 31 ($t \sim 210^{\rm d}$), 
but the width reported was narrower 
(FWHM $\sim 3,400$ km s$^{-1}$: Soderberg et al. 2005). 
This feature is marginally detected in our February spectrum, 
but the shape is uncertain because of the low S/N. 

We believe this is the H$_{\alpha}$ emission. 
Excessive emission at the red wing of [OI] $\lambda\lambda$6300, 6363 is 
sometimes observed in SNe Ib/c at relatively early epochs ($t \sim 100^{\rm d}$), 
and in such a case possible interpretation suggested to date 
is either [SiI] $\lambda$6527 
(e.g., 1997ef: Mazzali et al. 2004) or H$_{\alpha}$ 
(e.g., 1991A: Fillipenko 1991, see also Matheson 2001). 
The detection of this feature at $t \gsim 200^{\rm d}$ 
is not common, but there is at least one another SN showing a similar feature 
(SN 2004gn, which will be reported elsewhere). 

The feature, assuming it is H$_{\alpha}$, 
is either consistent with an emitting sphere 
with the outer boundary at $v \sim 10,000$ km s$^{-1}$ or 
an emitting shell bound between $v \sim 5,000 - 10,000$ km s$^{-1}$. 
The velocity at the outer boundary 
of the emitting H$_{\alpha}$ 
is similar to, but smaller than, 
the velocity of H ($v \sim 13,000$ km s$^{-1}$) seen in 
the spectrum at 2005 Apr 13 
($t \sim 15^{\rm d}$: Anupama et al. 2005; Tominaga et al. 2005). 
Since this velocity is very large as compared with 
the center of the $^{56}$Ni distribution along the line of sight 
($v \sim 2,000$ km s$^{-1}$: see \S\S 3.2 \& 3.3), 
it is probably difficult to ionize/excite H 
by the radioactive gamma-rays and resulting UV photons. 
More likely, the H$_{\alpha}$ comes from the ejecta decelerated by 
the weak CSM interaction. 
In any case, the detection of the high-velocity H$_{\alpha}$ supports 
the existence of 
the thin H envelope suggested by Anupama et al. (2005) and 
Tominaga et al. (2005). 
The line center of the H$_{\alpha}$ emission is consistent 
with the rest wavelength (Fig. 1), but the strong contamination 
in the blue wing by the [OI] makes the judgment difficult.

No strong emission is seen at [OI] $\lambda 5577$.  
The line ratio L([OI] $\lambda\lambda 6300, 6363$: $^{1}$D$_{2}$ $\to$ 
$^{3}$P)/L([OI] $\lambda 5577$: $^{1}$S$_{0}$ $\to$ $^{1}$D$_{2}$) 
is related to the electron number density ($n_{\rm e}$ [cm$^{-3}$]) and the 
electron temperature ($T_{3} \equiv T_{\rm e}/1000 {\rm K}$) as follows 
(under the usual assumption that the $^{1}$D$_{2}$ and $^{1}$S$_{0}$ levels 
are populated by thermal electron collisions). 

\begin{equation}
\frac{L_{6300+6363}}{L_{5577}} = 7.2 \beta_{6300} 
\frac{1 + 6.6 \times 10^{-9} n_{\rm e} T_{3}^{0.02}}
   {1 + 1.6 \times 10^{-6} n_{\rm e} T_{3}^{0.03}} 
   e^{\frac{25.83}{T_{3}}} \ ,
\end{equation}
where $\beta_{6300}$ is the Sobolev escape probability 
of the [OI] $\lambda 6300$ and about unity at the 
epoch of interest in the present paper.  
The expression is derived 
by solving rate equations for a simplified OI atomic model. 
It is correct to the first order in $n_{\rm e}$ and in the exponential 
term for $T_{\rm e}$. The form is somewhat different from 
that in Houck \& Fransson (1996), but these two expressions are 
consistent with each other in the density and temperature ranges 
of interest here ($n_{\rm e} \sim 10^{6} - 10^{10}$ cm$^{-3}$ 
and $T_{3} \sim 1 - 10$). 
Taking the rough estimate 
L([OI] $\lambda\lambda 6300, 6363$)/L([OI] $\lambda 5577$)
$\gsim 10$, the emitting region should be at relatively 
low temperature ($T_3 \lsim 4$, for the high density limit) 
and/or at low electron density ($n_{\rm e} \lsim 5 \times 10^6$ cm$^{-3}$, 
if $T_{3} = 10$).

The low density is supported from the 
large ratio of [CaII] $\lambda$7300 to CaII IR. 
The OI $\lambda$7774 is weak, further supporting the low density. 
It also suggests that ionization is low. 
SN 2005bf at $t \sim 300^{\rm d}$ belongs to the low density end of 
a typical condition seen in SNe Ibc at nebular phases 
with $n_{\rm e} \sim 10^{6} - 10^9$ cm$^{-3}$ 
(see, e.g., Fransson \& Chavalier 1989).

Figure 3 shows comparison of the December 26 spectrum 
with the spectra of SNe 1998bw and 1990I at similar epochs
\footnote{The spectrum of SN 1990I is taken from the SUSPECT 
(The Online Supernova Spectrum Archive) web page at 
http://bruford.nhn.ou.edu/\~\ suspect/index1.html, 
by courtesy of Abouazza Elmhamdi.}. 
Note that the flux is arbitrarily shifted for 
SNe 1998bw and 1990I for presentation. 
Despite the large difference in the luminosity (Fig. 2) and 
possibly in the line shapes and some line ratios, 
the overall features look similar among these objects. 
The [OI] $\lambda$6300/[CaII] $\lambda$7300 ratio in SN 2005bf 
is smaller than that of SN 1998bw. 
The oxygen core mass increases very sensitively as a 
function of $M_{\rm ms}$, while the explosively synthesized 
Ca does not. 
The smaller [OI]/[CaII] ratio thus indicates that the 
progenitor of SN 2005bf is less massive than 
SN 1998bw, i.e., $M_{\rm ms} < 40\Msun$. 
(See, e.g., Nakamura et al. 2001b and 
Nomoto et al. 2006 for the supernova 
yields. See Fransson \& Chevalier 
1987, 1989 for the theoretical [OI]/[CaII] ratios 
for the specific cases of $15\Msun$ and $25\Msun$ progenitor models. 
See also Maeda et al. 2007 for discussion on the 
[OI]/[CaII] ratio in other SNe Ib/c.) 

\subsection{Line Profiles and Blueshift}

\begin{figure}[tb]
\begin{center}
	\begin{minipage}[t]{0.4\textwidth}
		\epsscale{1.0}
		\plotone{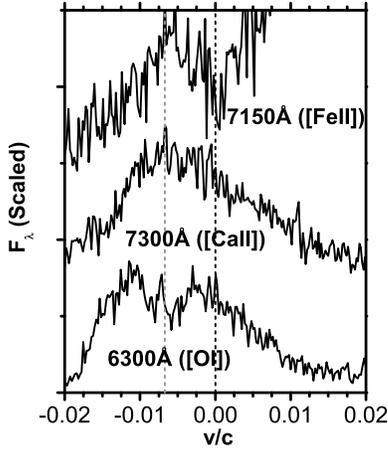}
	\end{minipage}
\end{center}
\caption[]
{Line profiles in velocity space (a minus sign for 
blueshift). The centers of the line are 
6300 ([OI]), 7150 (FeII), and 7300\AA ([CaII]). 
The velocity toward us is marked for 0 km s$^{-1}$ 
(black dashed) and for 2,000 km s$^{-1}$ (gray dashed). 
\label{fig4}}
\end{figure}

The observed line profiles and positions are very unique. 
Figure 4 shows line profiles around 6300, 7150, and 7300\AA, 
which we attribute to [OI] $\lambda\lambda$6300, 6363, 
[FeII] $\lambda$7155, and 
[CaII] $\lambda\lambda$7291, 7324. 
The line profiles are shown in velocity relative to 
6300, 7150, 7300\AA\ (after correcting for the host's redshift). 
All these lines 
show similar amount of blueshift relative to the rest wavelength 
($\sim 1,500 - 2,000$ km s$^{-1}$). 

The doubly-peaked profile of the [OI] is especially unique. 
This is clearly seen in the December 
spectrum, and is consistent with the low S/N February spectrum. 
Also, this feature is seen in another spectrum at a similar 
epoch taken independently (M. Modjaz, private communication). 
Thus, this peculiar line shape should be real. 
Note this is different from 
the doubly-peaked [OI] profile seen in SN 2003jd (Mazzali et al. 2005). 
In the case of SN 2003jd, it was basically symmetric with respect 
to the rest wavelength (i.e., no velocity shift), 
so that it was most naturally 
interpreted as oxygen distributed in a disk viewed from 
the equator (Maeda et al. 2002). 

By comparing these three lines, 
we can obtain insight on the distribution 
of materials. Between the two peaks in the [OI] emission, 
the [CaII] emits strongly, and the [FeII] is even more narrowly 
centered. 
The simplest interpretation is that the elements have layered distribution, 
i.e., Fe at the center of the emitting region ($v \sim 2,000$ 
km s$^{-1}$), which is surrounded by Ca, then by O.

\subsection{Spectrum Synthesis}

\begin{deluxetable*}{cccccccccccc}[tb]
 \tabletypesize{\scriptsize}
 \tablecaption{Spectrum Models\tablenotemark{a}
 \label{tab:model_02}}
 \tablewidth{0pt}
 \tablehead{
   \colhead{Model}
 & \colhead{$M_{\rm ej neb}$}
 & \colhead{$V$\tablenotemark{b}}
 & \colhead{$dV$}
 & \colhead{C}
 & \colhead{O}
 & \colhead{Na}
 & \colhead{Ca}
 & \colhead{$^{56}$Ni\tablenotemark{c}} 
 & \colhead{$T_{\rm e}$}
 & \colhead{log $n_{\rm e}$} 
 & \colhead{$\tau_{\rm abs}$\tablenotemark{d}}
}
\startdata
A  & 0.12 & 3,500 & 1,800 & 0.023 & 0.07 & 3.6E-5 & 7.2E-5&
0.024 & 5,100 & 6.4 & 0\\
B  & 0.34 & 5,000 & 0     & 0.068 & 0.21  & 1.0E-4 & 2.0E-4 & 
0.056 & 5,200 & 6.3 & 2 (67\% absorbed)\\
\enddata
\tablenotetext{a}{Units are the following. Masses ($\Msun$), 
velocity (km s$^{-1}$), $T_{\rm e}$ (K), and $n_{\rm e}$ (cm$^{-3}$).}
\tablenotetext{b}{$V$ and $dV$ are the outer velocity relative to 
the center of mass and the velocity shift with respect to the SN rest, 
respectively.}
\tablenotetext{c}{The mass of $^{56}$Ni ($\Msun$) initially synthesized at the explosion, 
before the radioactive decay.}
\tablenotetext{d}{The assumed dust optical depth $\tau_{\rm abs} = \kappa_{\rm abs} 
\rho V t$, where $t$ is the time since the explosion.}
\end{deluxetable*}

\begin{figure*}[tb]
\begin{center}
	\begin{minipage}[t]{0.45\textwidth}
		\epsscale{1.0}
		\plotone{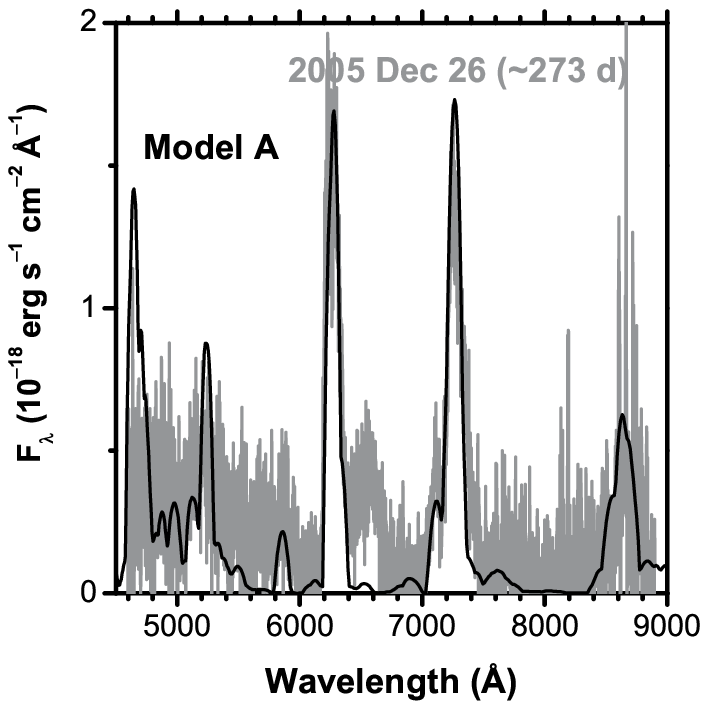}
	\end{minipage}
	\begin{minipage}[t]{0.45\textwidth}
		\epsscale{1.0}
		\plotone{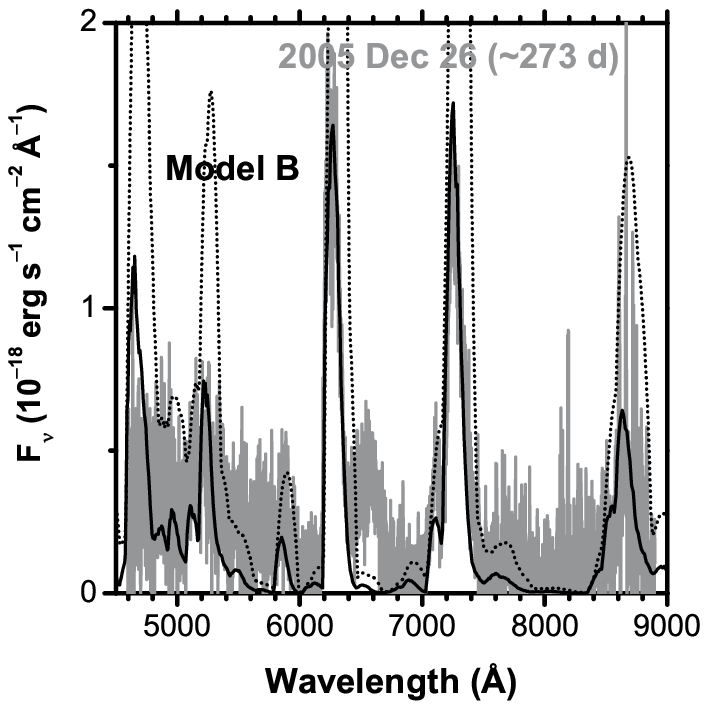}
	\end{minipage}
\end{center}
\caption[]
{Model spectra at $t = 270^{\rm d}$ (black solid), 
as compared with the December 26 spectrum (gray). 
For Model B, the original unabsorbed spectrum is also shown (dotted). 
See Table 2 for model parameters.  The distance modulus 
$\mu = 34.5$ and $E(B-V) = 0.045$ are adopted. 
\label{fig5}}
\end{figure*}

The similarity of the nebular spectra (Fig. 3) 
indicates that $n_{\rm e}$ and $T_{\rm e}$ are similar for 
SNe 2005bf, 1998bw, and 1990I at similar epochs. 
We have performed one-zone nebular 
spectrum synthesis computations. 
Following the $^{56}$Ni $\to$ $^{56}$Co $\to$ $^{56}$Fe 
decay chain as a heating source, the code computes 
gamma-ray deposition in a uniform nebula by the Monte-Carlo 
radiation transport Method. Positrons from the decays are assumed to be trapped  
completely (see \S 4.1). Positrons become a predominant heating source 
after the optical depth to the gamma-rays drops below $\sim 0.035$, 
following the density decrease. 
Ionization and NLTE thermal 
balance are solved according to the prescription 
given by Ruiz-Lapuente \& Lucy (1992). 
See Mazzali et al. (2001) and Maeda et al. (2006a) for details. 
Hereafter, we use the subscript "neb" for the values derived by 
modeling the nebular phase observations, i.e., 
$M_{\rm ej, neb}$, $M$($^{56}$Ni)$_{\rm neb}$, and so on 
(see Table 1). 

In the present models, we do not introduce He in the nebula. 
If the ejecta are heated totally by positrons, 
adding He does not affect the masses 
of the other elements derived in the spectrum synthesis. In this 
case, He has virtually nothing to do with both heating and cooling 
of the ejecta. On the other hand, if the ejecta are heated predominantly 
by gamma-rays, the situation is different. Increasing He mass fraction 
leads to lowering mass fractions of the other elements including 
$^{56}$Ni. However, to reproduce the observed total luminosity, 
reducing the fraction of $^{56}$Ni should be compensated by 
increasing the ejecta mass to absorb gamma-rays more effectively. 
Thus, the mass of the emitting materials ($M_{\rm ej, neb}$) derived 
without He is the lower limit. Likewise, 
the mass of each element, as well as that of $^{56}$Ni, obtained 
without He is the upper limit for the mass of each element, because the 
mass fraction for each element should be lower.

There are two possible ways to reproduce the blueshift in emission lines. 
One is the kinematical off-set in the distribution of the emitting 
materials (Model A: \S 3.3.1), and the other is the self-absorption within the ejecta 
reducing the contribution of light coming from the far side of the ejecta 
(Model B: \S 3.3.2). 
Figures 5 and 6 show the comparison between the model spectra and the December 
spectrum. Model parameters are listed in Table 2. 
Because of the one-zone treatment in the spectrum synthesis, we are not 
concerned with the detailed line profiles.

\begin{figure*}[tb]
\begin{center}
	\begin{minipage}[t]{0.3\textwidth}
		\epsscale{1.0}
		\plotone{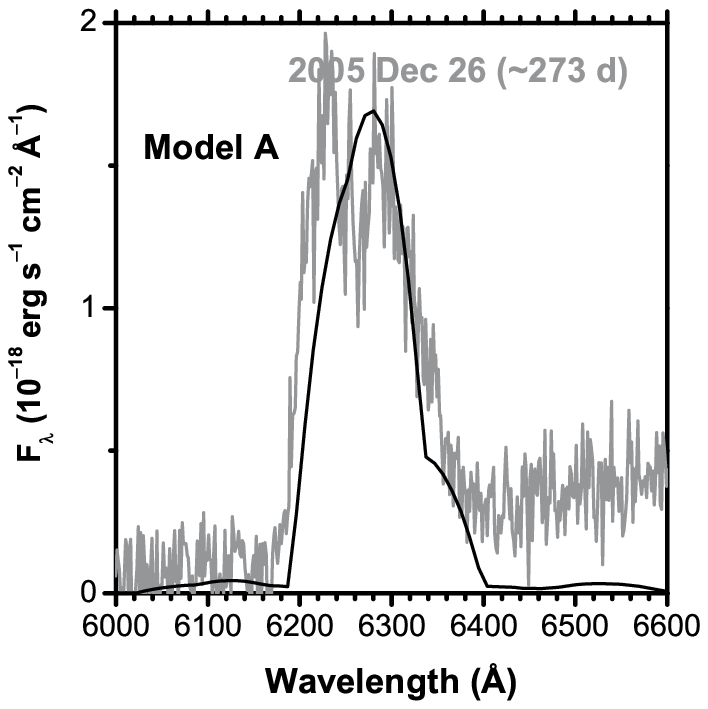}
	\end{minipage}
	\begin{minipage}[t]{0.3\textwidth}
		\epsscale{1.0}
		\plotone{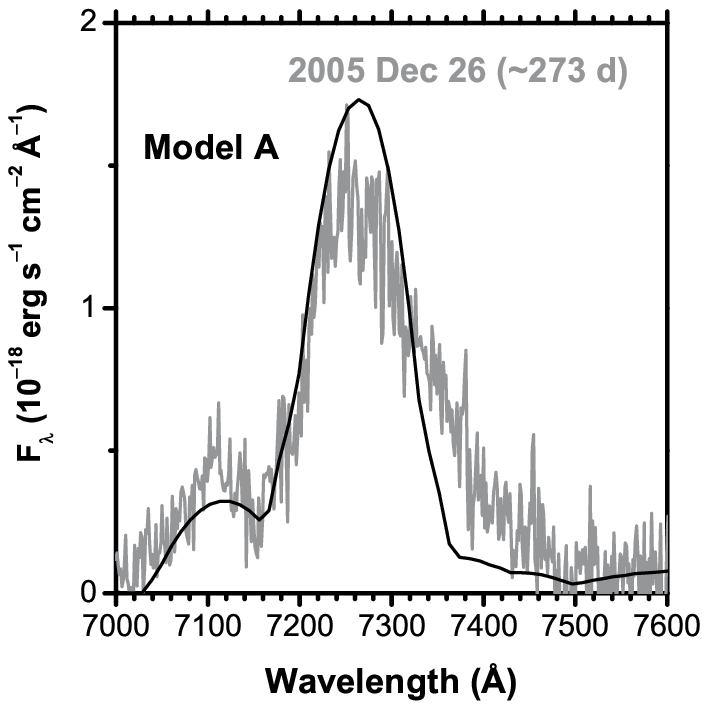}
	\end{minipage}\\
	\begin{minipage}[t]{0.3\textwidth}
		\epsscale{1.0}
		\plotone{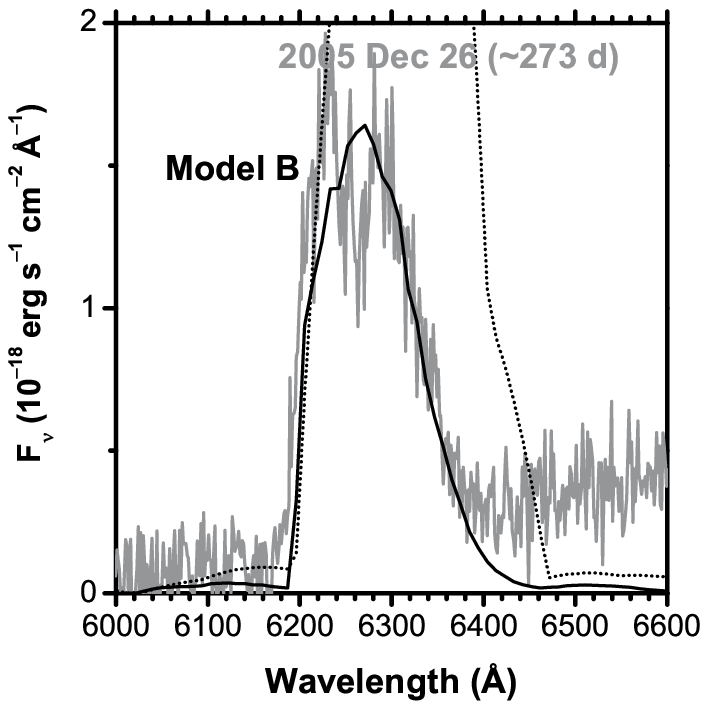}
	\end{minipage}
	\begin{minipage}[t]{0.3\textwidth}
		\epsscale{1.0}
		\plotone{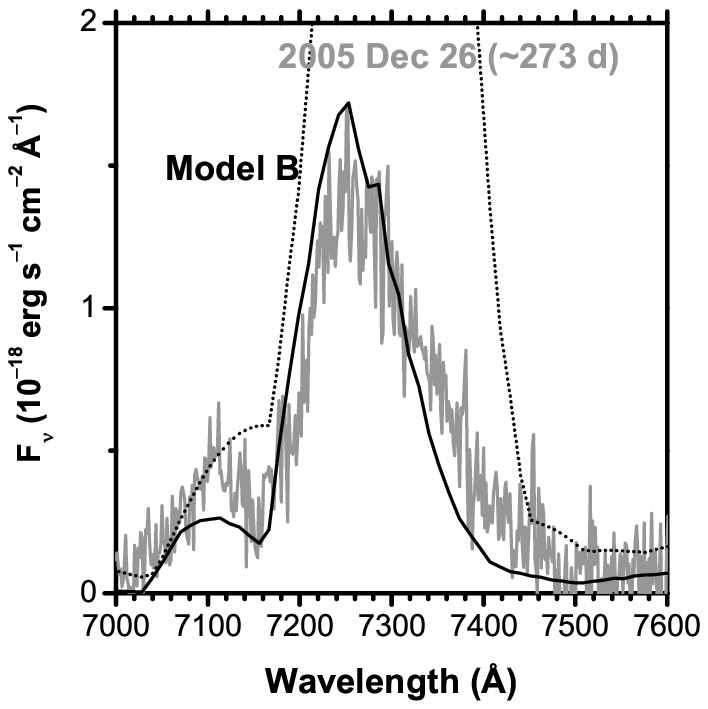}
	\end{minipage}
\end{center}
\caption[]
{Same with Figure 5, but for the [OI] $\lambda\lambda$6300, 6363 
and for the [CaII] $\lambda$7300 + [FeII] $\lambda$7150.  
\label{fig6}}
\end{figure*}

\subsubsection{Unipolar Blob: Model A}

In Model A, 
we neglect the optical radiation transport effect 
in the nebula, except for the optical depth effect within 
a line in the Sobolev approximation. 
As the observed spectrum shows blueshift 
relative to the expected line positions (Fig. 4: see also \S 3.2), 
we artificially shift the model spectrum blueward 
by $1,800$ km s$^{-1}$. The blueshift in this model 
is totally attributed to the kinematical distribution of the 
emitting materials (see e.g., Motohara et al. 2006). 
This is discussed later in this section. 

In Model A, $M_{\rm ej, neb} \sim 0.12\Msun$ 
and $M$($^{56}$Ni)$_{\rm neb} \sim 0.024\Msun$. 
These values are only $\sim 2\%$ and $8\%$ of 
$M_{\rm ej, peak}$ and $M$($^{56}$Ni)$_{\rm peak}$, respectively. 
Contrary, the same fractions are 
$\sim 25 - 50$\% ($M_{\rm ej}$) and $\sim 100$\% ($M$($^{56}$Ni)) 
for SN 1998bw (Mazzali et al. 2001).  
These values in Mazzali et al. (2001) are consistent with 
the expectation that in late phases we look into the $^{56}$Ni-rich 
region. In this sense, 
$M_{\rm ej, neb}$ and $M$($^{56}$Ni)$_{\rm neb}$ for SN 2005bf 
are too small to be compared with $M_{\rm ej, peak}$ and $M$($^{56}$Ni)$_{\rm peak}$. 

The electron density 
derived for SN 2005bf is similar to that for SN 1998bw at similar epochs 
(Mazzali et al. 2001). 
We find that introducing clumpy structures in SN 2005bf does not help. 
If the filling factor is smaller, then the oxygen mass 
should be even smaller to fit the [OI]$\lambda6300$ luminosity. 
Derived $n_{\rm e}$ (Table 2) is close to the critical density 
for the [OI]$\lambda6300$ emission, thus increasing 
$n_{\rm e}$ results in increasing the line emissivity 
per neutral oxygen. 

The CaII IR profile suggests a strong contribution from 
[CI] $\lambda$8727 in the red. 
If this is true, then we need relatively large mass ratio $\sim 0.35$ 
between C and O. 
This is consistent with the ratio for $M_{\rm ms} \lsim 20\Msun$ 
(e.g., Nomoto et al. 2006). 

Now we turn to the detailed element distribution. 
Figure 7 shows toy models to fit the line profiles, computed 
by assuming that 
the flux density is simply proportional to the density of 
homogeneous matter, 
and by artificially shifting the flux. 
As long as only the line profiles are concerned, 
various geometry can reproduce the observation. 
The blue shift suggests that 
a blob (or a jet) of $^{56}$Ni is 
ejected, and its center-of-mass velocity is $v \gsim 2,000$ km s$^{-1}$.  
Also, more centrally (but off-set from the SN rest) concentrated 
distribution of heavier elements yields a good fit to 
their narrower line profiles. 
If the viewing direction ($\theta$ measured from the pole) is close to the pole 
(Case A1: $\theta \sim 15^{\rm o}$), 
then the distribution of the oxygen should be more 
elongated to the same direction to explain the doubly-peaked [OI]. 
If $\theta$ is large (Case A2: $\theta \sim 75^{\rm o}$), on the other hand, 
the torus-like structure of oxygen-rich materials is necessary 
(Maeda et al. 2002; Mazzali et al. 2005). 
It should be interesting to examine in the future if 
these distributions can be reproduced by unipolar supernova explosion 
models (Hungerford, Fryer, \& Rockefeller 2005). 

In sum, in Model A, 
the spectrum of SN 2005bf at $t \sim 270^{\rm d}$ is explained 
by the ejection of a blob with $M_{\rm ej, neb}\sim 0.12\Msun$ and 
$M$($^{56}$Ni)$_{\rm neb}$ $\sim 0.024\Msun$.  
The blob is centered at $v \gsim 2,000$ km s$^{-1}$, distributed in 
$v \sim - 2,000 - \sim 5,000$ km s$^{-1}$ (Case A1) or 
$v \sim 0 - \sim 8,000$ km s$^{-1}$ (Case A2) depending on $\theta$ (Fig. 7).

\begin{figure*}[tb]
\begin{center}
	\begin{minipage}[t]{0.3\textwidth}
		\epsscale{1.0}
		\plotone{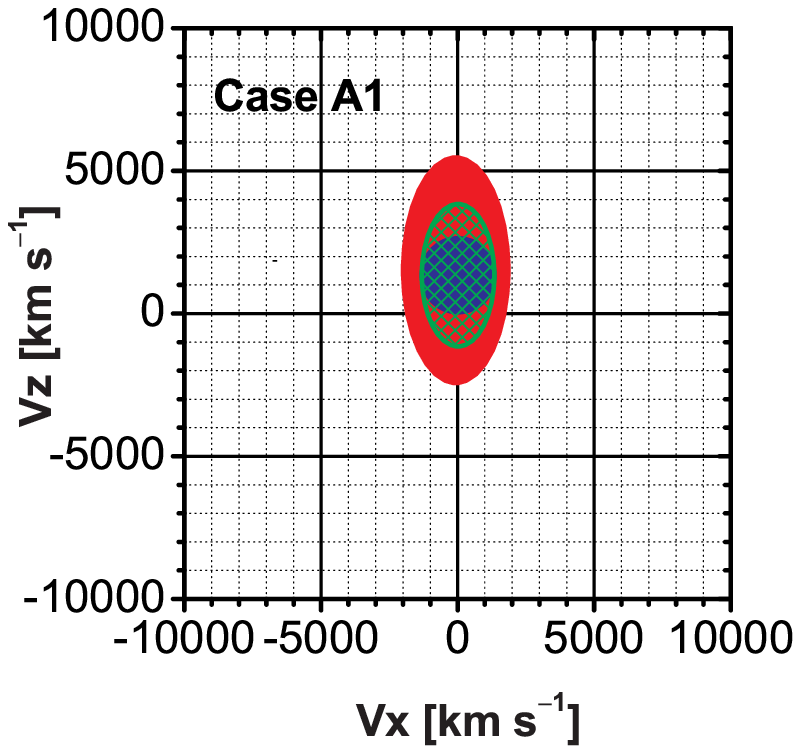}
	\end{minipage}
	\begin{minipage}[t]{0.3\textwidth}
		\epsscale{1.0}
		\plotone{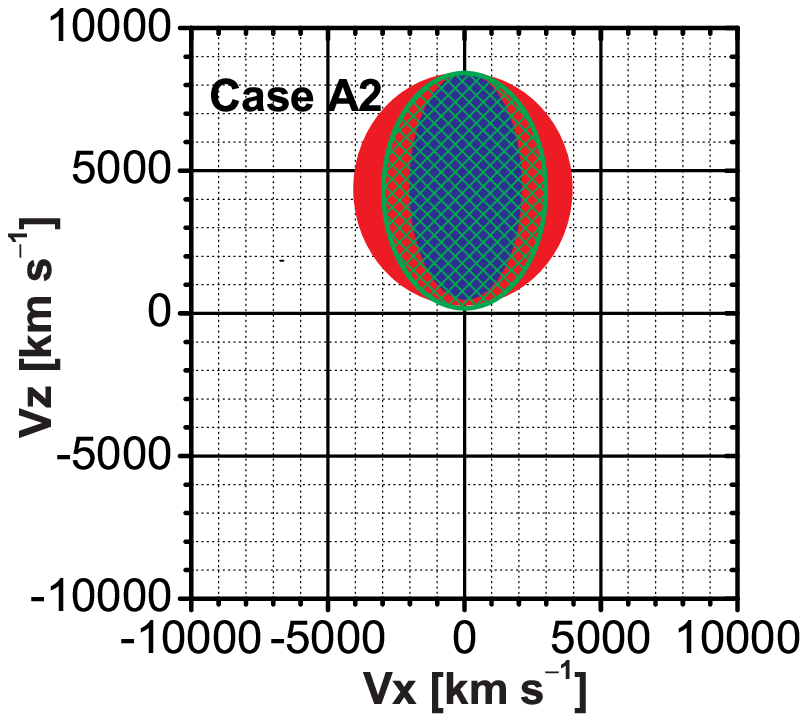}
	\end{minipage}\\
	\begin{minipage}[t]{0.3\textwidth}
		\epsscale{1.0}
		\plotone{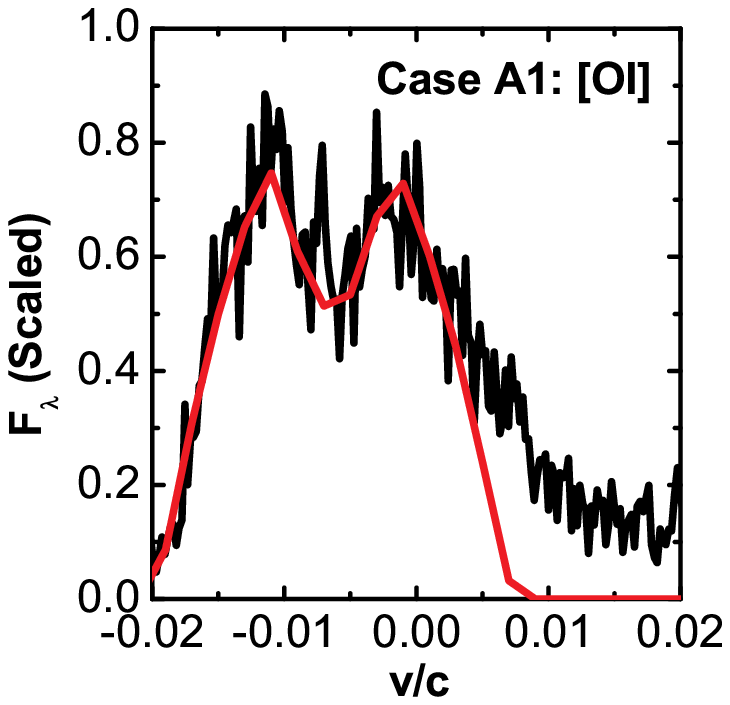}
	\end{minipage}
	\begin{minipage}[t]{0.3\textwidth}
		\epsscale{1.0}
		\plotone{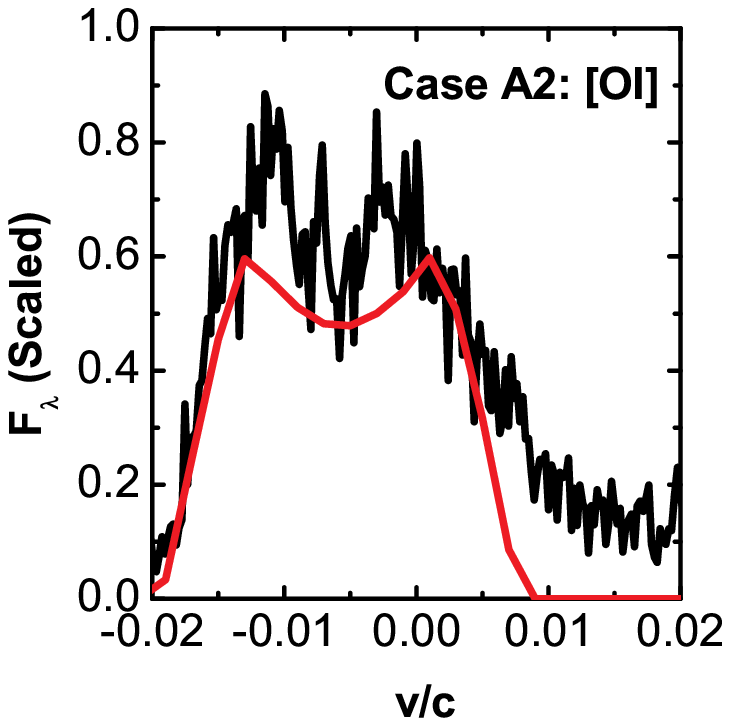}
	\end{minipage}\\
	\begin{minipage}[t]{0.3\textwidth}
		\epsscale{1.0}
		\plotone{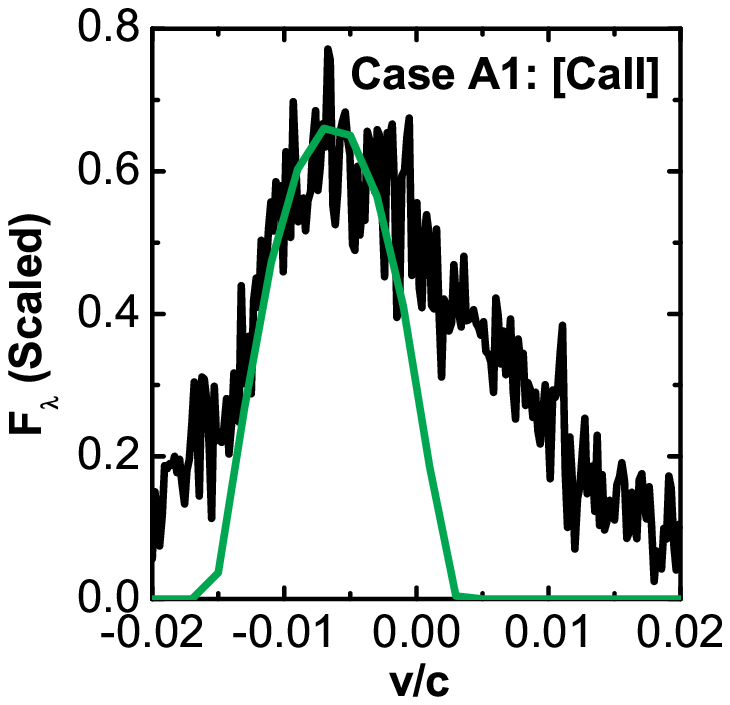}
	\end{minipage}
	\begin{minipage}[t]{0.3\textwidth}
		\epsscale{1.0}
		\plotone{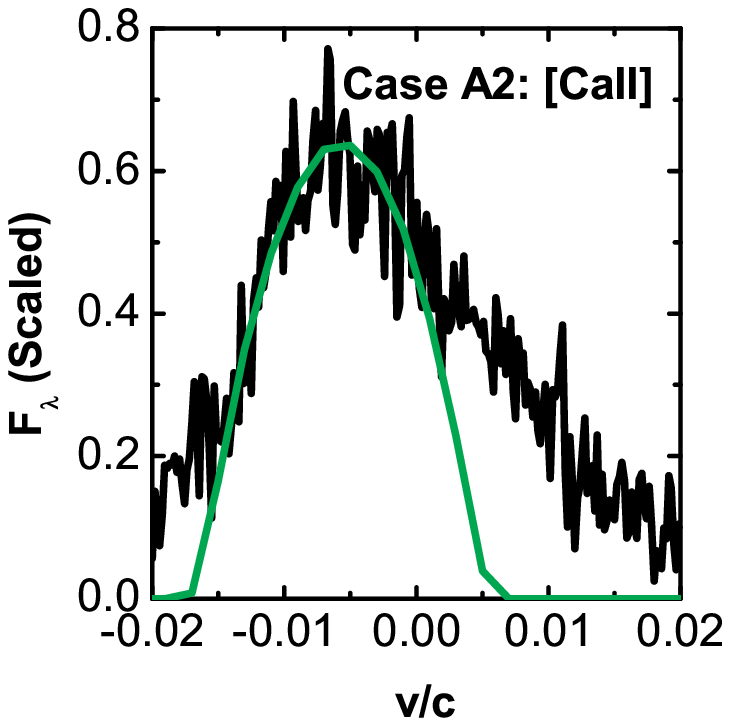}
	\end{minipage}\\
	\begin{minipage}[t]{0.3\textwidth}
		\epsscale{1.0}
		\plotone{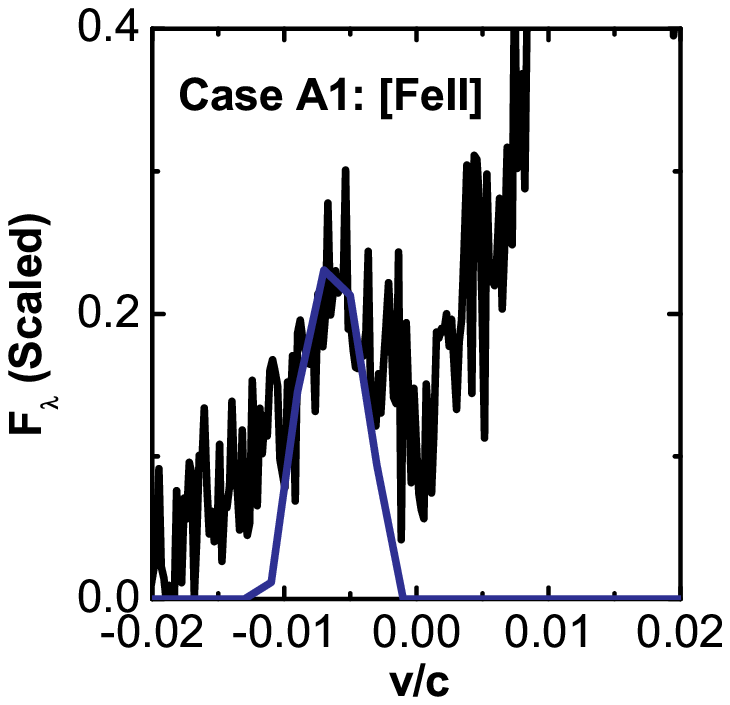}
	\end{minipage}
	\begin{minipage}[t]{0.3\textwidth}
		\epsscale{1.0}
		\plotone{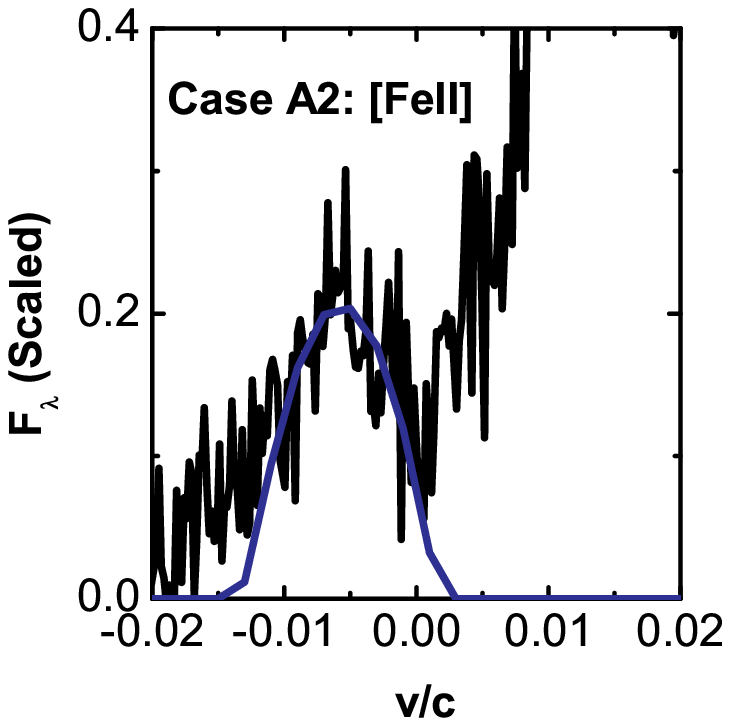}
	\end{minipage}
\end{center}
\caption[]
{Simple model fits to the line profiles. 
(a, Case A1: left panels) The blob model viewed at $\theta \sim 15^{\rm o}$ from the pole 
(i.e., $+ V_{z}$-direction). 
The distribution is shown for O (red), Ca (green), and Fe (blue). 
The expected line profiles are shown for the [OI] $\lambda\lambda$6300, 6363 (red), 
[CaII] $\lambda\lambda$7291, 7324 (green), and [FeII] $\lambda$7155 (blue) (from 
top to bottom). (b, Case A2: right panels) The blob model viewed at 
$\theta \sim 75^{\rm o}$ from the pole 
(i.e., $+ V_{x}$-direction). 
\label{fig7}}
\end{figure*}

\subsubsection{Self-Absorption: Model B}

Another interpretation is also possible for the blueshift of the 
emission lines. 
Figures 2 and 3 show 
the similarity between SNe 2005bf and 1990I 
in the light curve shape except for the peak at $t \sim 40^{\rm d}$, 
and in the nebular spectra. 
The early phase spectra could also be similar 
(A. Elmhamdi, private communication). 
The similarity may suggest that similar physical conditions 
could apply for these SNe. 

SN 1990I experienced the onset of 
blueshift in emission lines and accelerated fading in optical luminosity 
almost simultaneously (e.g., Elmhamdi et al. 2004). 
These are interpreted as the onset of dust formation and 
the self-absorption of optical light by the dust particles. 
Elmhamdi et al. (2004) constrained the fraction of the absorbed optical light 
$\sim 50\%$ at $t \sim 300^{\rm d}$ for SN 1990I. 

In Model B, we assume that the similar fraction of the optical 
light experiences  the absorption within the ejecta. We take the absorbed 
fraction to be $\sim 70\%$ (Table 2). 
In the spectrum synthesis for Model B, 
the optical light is dimmed as 
\begin{equation} 
I = I_{0} \exp(-\kappa_{\rm abs} \rho l) \ , 
\end{equation} 
where $I_{0}$ is the original intensity 
without absorption, and $l$ is the path length for each photon until escaping the 
nebula. The absorption opacity $\kappa_{\rm abs}$ 
is taken to be constant through the ejecta. The value of $\kappa_{\rm abs}$ 
is set by the requirement that the emergent luminosity is $\sim 30\%$ of the 
original luminosity.  

For the larger amount of absorption, the original 
luminosity should be larger to reproduce the observed luminosity. 
Accordingly, $M_{\rm ej, neb}$ and the mass of each element 
are larger in Model B than in Model A, as seen in Table 2. 

At the same time, the absorption dilutes the optical light from the far side 
selectively, thus causing the blueshift of the emission lines. 
Figure 6 shows that the blueshift similar to the observed one 
can be obtained although no kinematical off-set is assumed in Model B. 
The synthetic spectrum is bluer than observed for the [FeII] $\lambda$7150 and 
[CaII] $\lambda$7300, indicating that these elements are more centrally concentrated 
than oxygen, as is required in Model A. The absorption in the uniform sphere 
does not itself reproduce the doubly-peaked [OI] $\lambda\lambda$6300, 6363 doublet. 
The distribution of oxygen should be as shown in Figure 7, except for the 
center of the distribution which should be at the zero-velocity in Model B. 

In Model B, 
$M_{\rm ej, neb} \sim 0.34\Msun$ and $M$($^{56}$Ni)$_{\rm neb}$ $\sim 0.056\Msun$. 
The velocity of the outer edge of the emitting blob is 
$v \sim 5,000$ km s$^{-1}$, 
which is similar to that in Model A. 

\section{LATE TIME LIGHT CURVE}

\subsection{General Remarks}

The magnitude difference between the second peak 
($t \sim 40^{\rm d}$) and the late epoch 
($t \sim 270^{\rm d}$) is $\Delta B > 8.9$ mag 
and $\Delta R \sim 8.1$ mag.  
These are at least 2 magnitudes larger 
than seen in SNe 1998bw and 1990I (Fig. 2). 
Since the peak-to-tail luminosity difference is 
similar for different SNe Ib/c (e.g., Patat et la. 2001; 
Elmhamdi et al. 2004; Tomita et al. 2006; 
Richardson, Branch, \& Baron 2006), 
the very large difference in SN 2005bf is 
a unique property. 

\begin{figure}[tb]
\begin{center}
	\begin{minipage}[t]{0.4\textwidth}
		\epsscale{1.0}
		\plotone{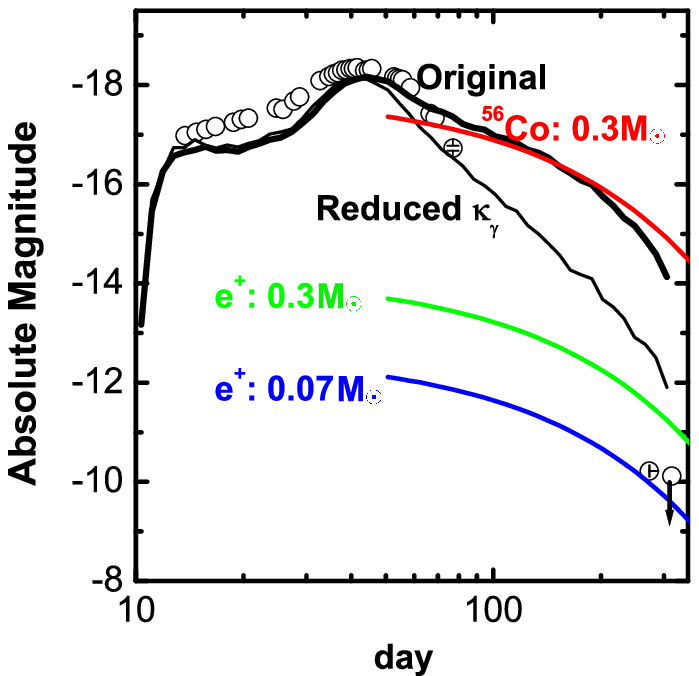}
	\end{minipage}
\end{center}
\caption[]
{The $R$-band light curve of SN 2005bf (circles), as compared with the model 
curves (black curves). The ejecta model is adopted from Tominaga et al. (2005), 
in which $M_{\rm ej, peak} = 7.3\Msun$, $E_{{\rm peak}, 51} = 1.3$, and 
$M$($^{56}$Ni)$_{\rm peak}$ $= 0.32\Msun$.  
Two models are shown (without/with reducing gamma ray opacity shown by 
thick/thin solid curve). 
Also shown is the energy generation rate by gamma-rays and positrons 
from $0.3\Msun$ of $^{56}$Co (red), 
and the energy generation rate only by positrons from $0.3\Msun$ 
(green) and $0.07\Msun$ (blue) of $^{56}$Co. 
\label{fig8}}
\end{figure}

Figure 8 shows the comparison between the $R$-band 
light curve of SN 2005bf 
and synthetic bolometric curves computed using the ejecta model 
of Tominaga et al. (2005) with $M_{\rm ej, peak} = 7.3\Msun$,  
$E_{{\rm peak}, 51}  = 1.3$ and $M$($^{56}$Ni)$_{\rm peak}$ $ = 0.32\Msun$. 
The light curves are computed by 
the Monte Carlo Radiation transport code described in 
Maeda et al. (2003) (see also Cappellaro et al. 1997). 
We adopt the absorptive opacity $\kappa_{\gamma} = 0.025$ cm$^{2}$ g$^{-1}$ 
for the gray gamma ray transport (see e.g., Maeda 2006c). 
The optical opacity prescription 
is similar to Tominaga et al. (2005): we assume that contribution from 
electron scatterings is equal to that from the line opacity 
(for the prescription for the line opacity, see Tominaga et al. 2005). 
This is largely consistent with 
the electron scattering opacity found in Tominaga et al. (2005) 
within a factor of two. 
Figure 8 shows that our synthetic light curve is consistent with 
the model curve computed by Tominaga et al. (2005). 
Also shown is the synthetic light curve computed using the reduced 
gamma ray opacity ($\kappa_{\gamma} = 0.001$ cm$^{2}$ g$^{-1}$) at $v < 5,400$ km s$^{-1}$ 
as examined in Tominaga et al. (2005). 

The $R$-band magnitude at $t \sim 270^{\rm d}$ 
is fainter than the bolometric magnitude 
expected from the model of Tominaga et al. (2005) by $\sim 5$  
mag (!) for the "brighter" expectation (without reducing 
$\kappa_{\gamma}$) or $\sim 3$ mag even for the "fainter" 
expectation (with reducing $\kappa_{\gamma}$). 
Since the $R$ magnitude is usually a good tracer of the bolometric magnitude in 
SNe Ib/c (see Fig. 2 for SN 1998bw), this large 
discrepancy is very odd and difficult to understand. 

Such a very rapid fading 
has never been observed in SNe Ib/c. 
The light curve of typical SNe Ib/c is reproduced 
by the energy input from gamma-rays 
and positrons produced in the 
radioactive decay chain $^{56}$Ni $\to$ $^{56}$Co 
$\to$ $^{56}$Fe. 
However, the newly observed light curve points of 
SN 2005bf turn out to be difficult to fit into this context. 
In late phases, the bolometric luminosity is equal to 
the energy of gamma-rays absorbed in the ejecta per unit time, 
as the radiation transfer effect is negligibly small. 
The gamma-ray optical depth can be estimated by 
\begin{equation} 
\tau_{\gamma} \sim 
1000 \times \frac{(M_{\rm ej}/\Msun)^{2}}{E_{51}} \times \left(\frac{t}{{\rm day}}\right)^{-2} 
\end{equation} 
(see e.g., Maeda et al. 2003). 
The model of Tominaga et al. (2005) or Folatelli et al. (2006) 
predict $\tau_{\gamma} \sim 0.5$ at $t = 300^{\rm d}$. For 
comparison, $\tau_{\gamma} \sim 0.02 - 0.06$ for SN 1998bw 
at $t = 300^{\rm d}$ 
($M_{\rm ej} \sim 10\Msun$, $E_{51} \sim 20 - 50$: 
Maeda, Mazzali, \& Nomoto 2006b; Nakamura et al. 2001a). 
Then, the peak-to-tail magnitude 
difference must be smaller in SN 2005bf than SNe 1998bw, which is 
inconsistent with what we have observed. 

Even more problematic is the fact that SN 2005bf in the late phase 
is even fainter than the lower limit set by the $^{56}$Co heating model. 
Positrons emitted from the $^{56}$Co decay 
produce energy at a rate $L_{e^+}$: 
\begin{equation}
L_{e^+} = 4.8 \times 10^{41} 
\left(\frac{M(^{56}{\rm Ni})}{M_{\odot}}\right) 
\exp{\left({-\frac{t}{113 \ {\rm day}}}\right)} \  {\rm erg} \ {\rm s}^{-1} \ .
\end{equation}
The positrons' mean free path is on 
the order of the gyroradius $r_{\rm gyr}$, 
which is 
\begin{eqnarray}
r_{\rm gyr} & = & \frac{\sqrt{2 m_{e} K} c}{e B} \nonumber\\
& \sim & 3.4 \times 10^3 \ {\rm cm}  \sqrt{\frac{K}{1 \ {\rm MeV}}} 
\left(\frac{B_{\rm mag}}{1 \ {\rm gauss}}\right)^{-1} \ .
\end{eqnarray}
Here $m_{e}$, $e$, $K$ are the mass, charge, and energy of the positron, 
$B_{\rm mag}$ is the strength of the magnetic field, and $c$ is the speed of light 
(all expressed in CGS-Gauss unit). A typical radius of the emitting 
supernova nebula is $\sim 10^{15}$ cm at $t = 300^{\rm d}$ with 
$v \sim 3,000$ km s$^{-1}$. Since the positrons' mean free path is 
many orders of magnitude smaller than the size of the nebula, 
all the positrons can be assumed to be trapped in the ejecta 
(at least in the absence of 
well aligned magnetic fields: see Milne, The, \& Leising 2001). 
This sets the lower limit of the bolometric luminosity for 
given $M$($^{56}$Ni).

It is seen from Figure 8 that the $R$-band and $B$-band magnitudes 
at $t \sim 270^{\rm d}$ are fainter than $L_{e^+}$ expected from 
$M$($^{56}$Ni)$_{\rm peak}$ $\sim 0.3\Msun$. 
Actually, the $R$ magnitude, assuming this is equal to 
the bolometric magnitude, 
is consistent with positron energy input from $0.08 \Msun$ of $^{56}$Ni 
according to equation (4). 
Thus, we set the strict upper limit $M$($^{56}$Ni)$_{\rm neb}$ $\lsim 0.08\Msun$. 
Since this input 
power has nothing to do with the gamma ray transport, reducing 
$\kappa_{\gamma}$ (Tominaga et al. 2005; Folatelli et al. 2006) 
does not help solve the discrepancy.

\subsection{Contribution from the Blob}

\begin{figure}[tb]
\begin{center}
	\begin{minipage}[t]{0.5\textwidth}
		\epsscale{1.0}
		\plotone{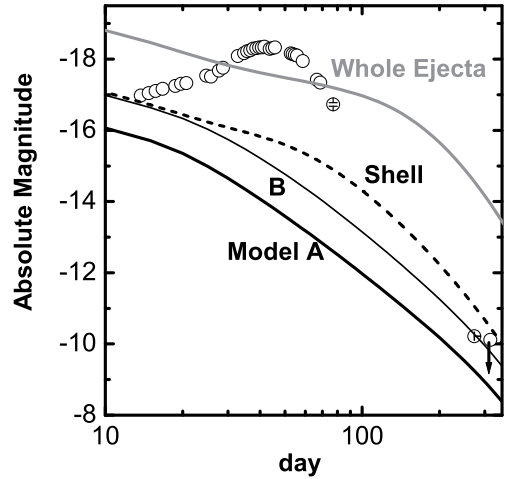}
	\end{minipage}
\end{center}
\caption[]
{Light Curve Analysis. The expected deposition luminosity is 
shown for the whole ejecta (gray solid; $M_{\rm ej, peak} = 7.3\Msun$, 
$E_{{\rm peak}, 51} =1.3$, and $M$($^{56}$Ni)$_{\rm peak}$ $= 0.3\Msun$), 
and for the $^{56}$Ni-rich shell 
(dashed; $M_{\rm ej} = 2\Msun$, $v = 3,900$ km s$^{-1}$, and 
$M$($^{56}$Ni) $= 0.06\Msun$) for the model of Tominaga et al. (2005). 
Also shown is the deposition luminosity expected for the blob 
derived from the nebular spectrum, i.e., models A (thick curve) and B 
(thin curve). 
\label{fig9}}
\end{figure}

As discussed in \S 3, the late phase spectra of SN 2005bf 
are dominated by the emission from a low mass blob with 
$M_{\rm ej, neb} \sim 0.1 - 0.4\Msun$ and 
$M$($^{56}$Ni)$_{\rm neb}$ $\sim 0.02 - 0.06\Msun$. 
The blob is either ejected with the central velocity 
$v \sim 2,000$ km s$^{-1}$ (if viewed from the pole) -- 
$\sim 5,000$ km s$^{-1}$ (if viewed $\sim 75^{\rm o}$ 
away from the pole), or it suffers from the absorption within the ejecta 
like SN 1990I. In either case, 
the emitting materials are distributed up to $v \sim 5,000 
- 8,000$ km s$^{-1}$ (\S 3).

A contribution of the blob component to the light curve is estimated in 
Figure 9. Taking $M_{\rm ej, neb}$, $E_{{\rm neb}, 51}$, and 
$M$($^{56}$Ni)$_{\rm neb}$ from Table 1, 
we use equations (1 -- 3) of Maeda et al. (2003) with the modification 
to include contribution from the $^{56}$Ni $\to$ $^{56}$Co decay. 
In Model B, the luminosity in the nebular epochs ($t \sim 300^{\rm d}$) should be 
further decreased by $\sim 70\%$ ($\sim 1$ mag) from the curve in Figure 9 
to take into account the self-absorption. 

It is seen that the energy deposition curve expected from this blob 
(both models A and B) roughly connects 
the first peak ($t \sim 20^{\rm d}$) and the late Subaru points 
($t \sim 300^{\rm d}$). 
That is to say, the optical output from SN 2005bf was 
dominated by this blob component in the earliest and the late 
epochs, while around the main peak ($t \sim 40^{\rm d}$) 
the emission from the whole ejecta 
made the predominant contribution.

Also shown in Figure 9 is the deposition curve expected from the 
$^{56}$Ni-rich shell 
($3,900$ km s$^{-1}$ $< v <$ $5,400$ km s$^{-1}$) of Tominaga et al. (2005). 
The shell has the mass $\sim 2\Msun$ in which $M$($^{56}$Ni) $\sim 0.06\Msun$. 
Interestingly, the curve expected from the shell is similar to the blob contribution 
derived by the nebular spectra. 
The similar amount of $^{56}$Ni and the similar velocity 
between the blob and their shell  
suggest that what Tominaga et al. (2005) attributed to the shell is actually 
the blob we derived in this study. 
However, the total mass of the blob ($M_{\rm ej, neb} \lsim 0.4\Msun$) 
is smaller than what they derived ($\sim 2\Msun$).  

Possibly, detailed (2D/3D) ejecta structure and/or 
the optical opacity prescription affect the mass estimate. 
These effects are more important in the early phase modeling, 
since in the early phases only a small fraction of the ejecta is seen and 
the optical transport effect is strong.

\section{DISCUSSION}

The late phase data presented in this paper 
add the following peculiarities to SN 2005bf. 
(1)
It is extremely faint at late phases. 
(2) 
Line emissions are blueshifted by $\sim 1,500 - 2,000$ km s$^{-1}$. 

The extreme faintness needs special condition (\S 4). 
To explain the faintness, 
there are at least four possibilities. 
\begin{itemize}
\item[(a)] The ejecta are much more transparent to 
gamma-rays and even to positrons than in other SNe. 
\item[(b)] The fraction of radiation output in the optical range is 
extremely small ($\sim 1$ \%). 
\item[(c)] $M$($^{56}$Ni) decreases with time, from $M$($^{56}$Ni)$_{\rm peak}$ 
$\sim 0.3\Msun$ at $t \sim 40^{\rm d}$ to $M$($^{56}$Ni)$_{\rm neb}$ 
$\lsim 0.1\Msun$ at $t \sim 270^{\rm d}$. 
\item[(d)] The peak ($t \sim 40^{\rm d}$), at least, was not powered by the $^{56}$Ni decay chain. 
\end{itemize}
In this section, we discuss these possible interpretations.

\subsection{Gamma-ray and Positron escape?} 

The drop of the light curve was already observed between 
$t \sim 40^{\rm d}$ and $\sim 80^{\rm d}$. 
It was suggested that it could be explained by assuming the 
reduced gamma-ray opacity $\kappa_{\gamma}$ 
(Tominaga et al. 2005; Folatelli et al. 2006). 
This is possible for the period 
$t \sim 40^{\rm d} - 80^{\rm d}$, 
but we show in the following that this 
is unlikely to work at the late epochs ($t \sim 300^{\rm d}$). 

They argued that the reduction of $\kappa_{\gamma}$ 
could take place if the geometry of the ejecta is far from spherically symmetric. 
If the ejecta have large inhomogeneity (clumpy or jet-disk structure) 
in the density structure, the effective optical depth is 
reduced as 
\begin{equation} 
\tau_{\rm eff} = \frac{\tau_{0}}{\tau_{\rm c}} (1 - \exp(-\tau_{\rm c})) 
\end{equation} 
(Bowyer and Field 1969; Nagase et al. 1986). 
Here $\tau_{0}$ and $\tau_{\rm c}$ are the gamma-ray optical depths for homogeneous medium 
with the same average density and for each dense structure (e.g., clump). 

We note that this effect works only when $\tau_{\rm c} \gg 1$. 
As long as the clumps (or dense regions) follow the homologous expansion, 
$\tau_{\rm c}$ should decrease with time according to $\tau_{\rm c} \propto t^{-2}$. 
It is then expected that this opacity 
reduction effect does not work in the nebular phases. Furthermore, 
the nebular spectra indicate that very dense regions such that $\tau_{\rm c} \gg 1$ 
to gamma-rays do not exist (\S 3). 
Thus, this effect can not be used as an argument for the reduction of $\kappa_{\gamma}$ 
in the nebular phase. 

Even worse, not only gamma-rays, but also positrons, should escape 
the ejecta effectively in this interpretation. It is even more difficult 
to explain an enhancement of the amount of positrons that escape the 
ejecta without interacting, since positrons have much smaller mean free 
path than gamma-rays (equation (5)). In sum, we conclude that 
the gamma-ray and positron escape scenario is unlikely.

\subsection{Absorption in the Ejecta?}

Qualitatively, the two features in the late phases 
(faintness and blueshift) could be expected from 
self-absorption in the SN ejecta. 
These features are essential in Model B to fit the December spectrum. 
Note that Model B yields a light curve shape similar to that of SN 1990I 
(see Figs 2 and 9). 
In this section, we consider a more extreme case 
than in Model B, 
and examine whether the self-absorption 
within the ejecta of $M_{\rm ej, peak} \sim 7\Msun$ 
and $M$($^{56}$Ni)$_{\rm peak}$ $\sim 0.3\Msun$ 
can explain the entire light curve of SN 2005bf. 

This is one possibility. 
However, the following arguments can be used against 
(although not definitely) the extreme self-absorption scenario. 
Some are related to the light curve shape. 
\begin{itemize}
\item[(1)] Figure 2 
shows that the light curve starts dropping faster than other SNe Ibc 
already at $t \sim 50^{\rm d}$, and 
this is likely the beginning of the very faint nature of SN 2005bf. 
Such a drop at a relatively early phase is seen neither in SN 1990I nor 
in other SNe undergoing dust formation. 
The temperature in the ejecta at $t \sim 50^{\rm d}$ should 
be too high to form dust in the ejecta (e.g., Nozawa et al. 2003). 
Observationally, NIR contribution is estimated to be 
$\sim 50$ \% at $t \sim 80^{\rm d}$, which is similar to 
SNe 2002ap and 1998bw (Tomita et al. 2006), indicating 
the temperature is similar to these objects.
\item[(2)] If the rapid fading of SN 2005bf is caused by the self-absorption, 
almost all the radiation must be emitted in NIR (Near Infrared) -- FIR 
(Far Infrared). Such an extreme absorption is not seen in the dust forming SNe. 
For example, the fraction of the absorbed emission is estimated to be 
$\sim 50\%$ at $t \sim 300^{\rm d}$ for SN 1990I (Elmhamdi et al. 2004). 
For SN 2005bf, unfortunately, no NIR or FIR observation at the late epochs 
is available. 
\end{itemize}

The other arguments are related to the spectral features. 
\begin{itemize}
\item[(3)] According to (2) above, most of the light in the optical should 
be absorbed, and thus only the bluest portion of each emission line 
should be observed in this scenario. 
We note, however, that the modest model B ($70\%$ absorption) 
is already consistent with the observed wavelength shift.  
\item[(4)] This scenario with almost 100 \% absorption would 
result in an extremely large [OI]$\lambda$ 6300/[CaII]$\lambda$ 7300 ratio, 
since oxygen (which is surrounding the Ca-rich region) is expected to suffer 
from less absorption (see (3) above). 
This is inconsistent with 
the observed ratio which is smaller than that in SN 1998bw (see \S 3.1 and 
Figure 3). For example, if we use the stratified model of 
Tominaga et al. (2005) and take the absorption fraction to be $95\%$, 
we find that the [CaII] line almost vanishes while the [OI] line 
is still brighter than observed by a factor of $\sim 5$.
\end{itemize}

In conclusion, the examination in this section suggests that the extreme 
self-absorption scenario is unlikely. However, 
we missed the most important information for the judgment, 
i.e., NIR to FIR observations at the nebular phases.

\subsection{Fallback?}

\begin{figure}[tb]
\begin{center}
	\begin{minipage}[t]{0.5\textwidth}
		\epsscale{1.0}
		\plotone{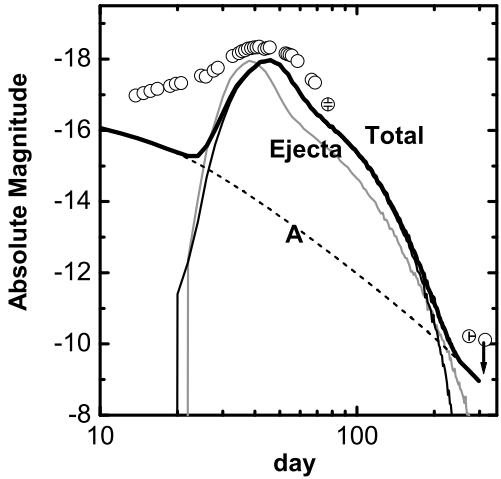}
	\end{minipage}
\end{center}
\caption[]
{Fallback model light curves. 
Shown here are the fall back model with 
$t_{\rm acc} = 40^{\rm d}$, $\dot M_{\rm ej} = 1.2 \times 10^{-2}\Msun {\rm d}^{-1}$ 
(black thin solid), and 
$t_{\rm acc} = 20^{\rm d}$, $\dot M_{\rm ej} = 2.0 \times 10^{-2}\Msun {\rm d}^{-1}$ 
(gray thin solid).  
The contribution from the blob (Model A) 
is also shown (dashed). 
The sum of the ejecta contribution (for the model with 
$t_{\rm acc} = 40^{\rm d}$) and the blob contribution is shown by the thick curve. 
See Appendix A for details. 
\label{fig10}}
\end{figure}

In the following scenarios (\S 5.3 and \S 5.4), 
it is interpreted that the low mass blob dominates the optical light 
in the first peak and the late phase (\S 4.2). 
We examine whether the second peak can be 
reproduced by any scenarios without producing too strong emission 
in the late phase. 
If this condition is satisfied, the entire light curve 
could be explained by the combination of the second 
peak component plus the blob component. 

We use the ejecta model of Tominaga et al. (2005) with 
$M_{\rm ej, peak} = 7.3\Msun$ and $E_{{\rm peak}, 51} = 1.3$. 
As we have already replaced the "high velocity" $^{56}$Ni 
component of Tominaga et al. (2005) with the 
low mass blob (Model A or B; see \S 4.2), 
we set the mass fraction of 
$^{56}$Ni at $v > 1,600$ km s$^{-1}$ to be zero hereafter 
(the contribution of the blob is added to the synthetic light curve 
{\it after} the computation of this ejecta contribution). 

One possible process that decreases $M$($^{56}$Ni) with time 
is the fallback of 
the inner $^{56}$Ni-rich region onto the central remnant 
(e.g., Woosley \& Weaver 1995; Iwamoto et al. 2005).  
Figure 10 shows the examples of synthetic light curves 
of supernovae hypothetically undergoing fallback 
(see Appendix A for details).  

The model assumes that the accretion begins 
at a specific time ($t_{\rm acc}$), and that 
the mass accretion rate after the time $t_{\rm acc}$ obeys 
the form $\dot M_{\rm ej} \propto t^{-5/3}$. These are qualitatively 
expected in spherically symmetric fallback (e.g., 
Woosley \& Weaver 1995; Iwamoto et al. 2005). 
The light curve of SN 2005bf can be reproduced in this context 
{\it only if} we assume $t_{\rm acc} \gsim 20^{\rm d}$. 

Difficulties encountered in the spherical 
fallback scenario are the following (Fig. 11; 
see Appendix B for details): 
\begin{itemize} 
\item[(1)] $E_{{\rm peak}, 51} \sim 1 - 1.5$ 
is too large to cause the spherical hydrodynamic fallback. 
\item[(2)] Fallback should take place very late, later than 10 days since 
the explosion. The time scale of the spherical fallback in 
the He star progenitor model is much shorter than 10 days (Figure 11). 

\end{itemize}

The problem in the energy may be relaxed by 
considering an asymmetric explosion, 
because the velocity in the weak-explosion direction may be 
sufficiently small compared with escape velocity. 
The problem in the time scale may be overcome by 
introducing some mechanism to delay the fallback, e.g., 
by disk accretion (e.g., Mineshige, Nomoto, \& Shigeyama 1993). 
However, it seems difficult to realize the condition 
that the $^{56}$Ni-rich region experiences the fallback 
in a period of $\sim 1$ month (e.g., Figures 2 -- 4 of Mineshige et al. 1997). 

\begin{figure}[tb]
\begin{center}
	\begin{minipage}[t]{0.5\textwidth}
		\epsscale{1.0}
		\plotone{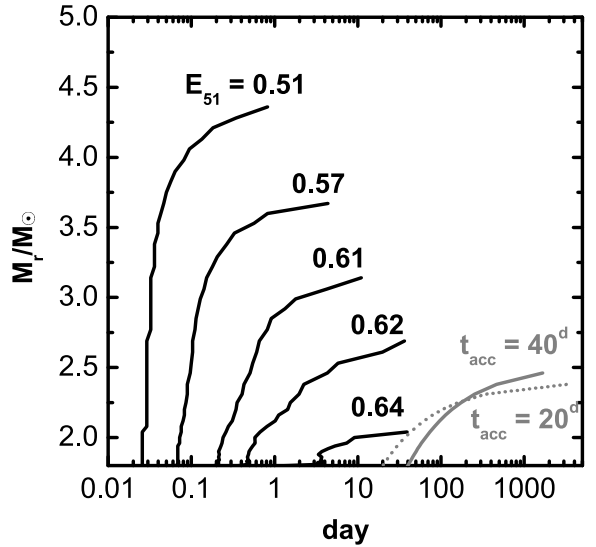}
	\end{minipage}
\end{center}
\caption[]
{The fallback mass as a function of time as obtained 
by a set of hydrodynamic simulations  
with varying explosion energy for a $7.3\Msun$ He star model 
(black curves; the values of the final kinetic energy 
are shown in the figure). Also shown are those used to reproduce the light curve 
(Fig. 10) with two different fallback time scale (gray curves). 
See Appendix B for details. 
\label{fig11}}
\end{figure}

\subsection{Central Remnant's Activity?}

Another possibility is a different type of the 
energy source for the second peak. 
Other than $^{56}$Ni and $^{56}$Co, 
a possible heating source is the interaction between 
the ejecta and CSM. However, there is a 
strong argument against its responsibility for energizing the 
light curve of SN 2005bf. The light curve shows 
a slow rise to the peak ($t \sim 40^{\rm d}$), which is typical 
characteristics of diffusion of photons from deep 
in the ejecta. Also, there is no indication of strong interaction 
in its spectra around the peak. 

The heating source should be buried deep in the ejecta, 
and it should be capable of producing the total 
energy input $\gsim 10^{49}$ erg, 
at the maximum rate of $\gsim 10^{43}$ erg s$^{-1}$. 
Except radioactivity, a possible source that satisfies these conditions 
could be the activity of the central compact remnant. 
Indeed, the peculiar features in the early phases led Tominaga et al. 
(2005) to speculate the formation of 
a strongly magnetized neutron star (a magnetar). 
Folatelli et al. (2006) speculated that SN 2005bf would be driven by 
the central engine similar to that in gamma-ray bursts, for which 
a popular idea is a black hole and an accretion disk system (Woosley 
1993). 

Since the luminosity emitted from a system consisting of a black hole 
and an accretion disk is expected not to exceed $\sim 10^{43}$ erg s$^{-1}$ 
because of neutrino losses (e.g., Janiuk et al. 2004), we consider a 
potentially more effective mechanism of emitting photons, i.e., 
a magnetar.  The energy input is assumed to take the following form 
as a function of the position in the ejecta ($v$, expressed 
in velocity space) and time ($t$): 
\begin{equation} 
L_{\gamma} (v, t) 
= L_{0} \left(1.0 + 2.0 \times \frac{t}{t_{0}}\right)^{-\beta} d(v) \ , 
\end{equation} 
where $d(v) = D \exp(-v/v_{0})$ 
if $v \le 3000$ km s$^{-1}$ and $d(v) = 0$ 
if $v > 3000 {\rm km s}^{-1}$, with $D$ the normalized constant. 
Here $L_{0}$ is the initial energy injection rate 
in the form of high energy photons, $t_0$ is characteristic time scale, 
and $v_{0}$ is characteristic length scale, and 
$\beta$ is the decay temporal index. 

We assume the photon index of $-2.5$, as is similar to that of the Crab Pulsar 
(e.g., Davidson \& Fesen 1985). 
We also assume that the minimum energy of the photon is 1 keV 
for the input high energy spectrum. 
Since the optical depth to these high energy photons is very large 
at the epochs considered here, details of the spectral index and the 
cut-off energy do not affect the result sensitively (see also 
Kumagai et al. 1991). 

The density distribution of the ejecta 
model is taken from Tominaga et al. (2005) 
with the reconstruction of $M_{\rm ej}$ and $E_{51}$ in a self-similar manner. 
We set $^{56}$Ni mass fraction zero throughout the ejecta to 
investigate the contribution of this hypothetical energy source. 
In the model shown in Figure 12, we used $M_{\rm ej} = 8.0\Msun$, $E_{51} = 1.3$, 
which is within the range to explain the early phase spectra 
(Tominaga et al. 2005). 

With the energy input and the ejecta model, 
the high energy radiation transport is solved by 
the Monte Carlo code described in Maeda (2006c). The optical photon 
transport is solved by the method described in \S 4 (see also Maeda et al. 2006b). 
Figure 12 shows the model with the following parameters: 
\begin{itemize}
\item $L_{0} = 8 \times 10^{43}$ erg s$^{-1}$, 
\item $t_{0} = 60^{\rm d}$, 
\item $v_0 = 2,500$ km s$^{-1}$, and 
\item $\beta = 4$. 
\end{itemize}

The position of the second peak is roughly reproduced irrespective of 
the input parameters, since the 
diffusion time scale mainly determines the peak date. 
The large peak luminosity and the rapid decline after the second peak 
can qualitatively be explained if $L_0 \gsim 10^{44}$ erg s$^{-1}$ 
and $t_{0} \lsim 60^{\rm d}$. 
Large $L_{0}$ is expected if the central remnant is 
a strongly magnetized neutron star. 
The relations between the model parameters and 
physical quantities are discussed in \S 6.

\begin{figure}[tb]
\begin{center}
	\begin{minipage}[t]{0.5\textwidth}
		\epsscale{1.0}
		\plotone{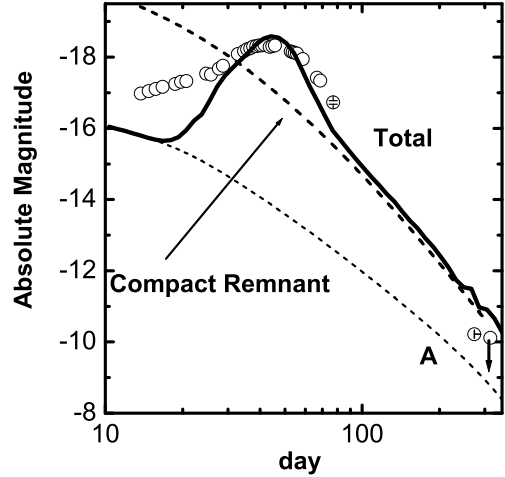}
	\end{minipage}
\end{center}
\caption[]
{Synthetic light curve with the putative energy input from the central remnant 
(thick solid).  
The model parameters are as follows: 
$L_{0} = 8 \times 10^{43}$ erg s$^{-1}$, $t_{0} = 60^{\rm d}$, $\beta = 4$. 
The input luminosity of the remnant (thick dashed) 
and the contribution from the blob (thin dashed) are also shown. 
See \S 5.4 for details. 
\label{fig12}}
\end{figure}

\section{The HYPOTHESIS -- A BIRTH EVENT OF A MAGNETAR}

In \S 5, we have shown that the light curve of SN 2005bf is 
explained by the energy input from a magnetar left behind 
the explosion. 
In this section, we relate the model parameters 
to physical quantities, and discuss consequences and 
implications of the scenario. 

\begin{itemize}

\item[(1)] $L_0 \gsim 10^{44}$ erg s$^{-1}$ 
and $t_{0} \lsim 60^{\rm d}$ is required to reproduce 
the large peak luminosity and the rapid decline after the second peak. 
If interpreted as a pulsar energy input, 
these two conditions are satisfied only if the remnant is a millisecond 
magnetar (i.e., the surface magnetic field is $B_{\rm mag} \sim 10^{14 - 15}$ gauss, 
and the initial spin period $P_0 \sim 5 - 10$ ms). 

The model parameters, 
$L_{0} = 8 \times 10^{43}$ erg s$^{-1}$ and $t_{0} = 60^{\rm d}$, 
corresponds to a pulsar with the magnetic field 
$B_{\rm mag} \sim 3 \times 10^{14} (P_0/10 \ {\rm ms})^{2} \sqrt{0.1/f_{\rm rad}}$ gauss, 
using the dipole radiation formula 
(Ostriker \& Gunn 1969; see also Rees \& Meszaros 2000). 
Here $P_0$ is the initial spin period and 
$f_{\rm rad}$ is a fraction of energy going into the radiation.  
The total energy injection with these parameters is $\sim 7 \times 10^{50}  
(f_{\rm rad}/0.1)^{-1}$ erg, a fraction of which 
might be consumed to increase the 
kinetic energy of the SN ejecta to $E_{51} = 1 - 1.5$ and/or 
to develop the pulsar nebula in the early phase (see below).

\item[(2)]
The relatively large breaking index $\beta = 4$ is required to reproduce 
the large contrast between the peak and the tail, but it is still within 
the range of the decay rates inferred for galactic pulsars. 

The temporal index $\beta = 2$ 
is expected for the energy input from a pulsar slowed down predominantly by 
the magnetic dipole radiation. It is also the case for similar models involving 
the conversion of the rotational energy to the energy of 
radiation or relativistic particles mediated purely by the magnetic field 
(e.g., Ostriker \& Gunn 1969). 

If $\beta = 4$, then 
the pulsar's breaking index is $n = 2$ 
($\dot \Omega \propto \Omega^{n}$ by definition, where $\Omega$ is the rotational 
angular frequency), while the magnetic dipole model ($\beta=2$) predicts $n = 3$. 
The breaking index as small as 2 is expected by dissipation processes 
mediated not only by the magnetic field (e.g., Menou, K., Perna, R., 
\& Hernquist 2001), 
and is really inferred for most pulsars with the index measurement available 
(e.g., Livingstone, Kapsi, \& Gavriil 2005 and references therein).

\item[(3)]
Although the peak date can be reproduced irrespective of $v_0$ 
as the diffusion time scale mainly determines the peak date, 
a good fit to the light curve width around the peak is obtained 
if we set $v_0 \sim 2,000 - 2,500$ km s$^{-1}$. 

This is larger than the average expansion velocity of the Crab pulsar 
wind nebula seen in X-rays (e.g., Mori et al. 2004):  
Assuming 2 kpc for the distance to the Crab nebula, 
the spatial extent of its X-ray image corresponds to 
the average expansion velocity of $\sim 500$ km s$^{-1}$. 
The early development and the relatively large size of 
the pulsar nebula in our light curve model might not be surprising, 
because of small $t_{0}$ and large $L_{0}$. 
Injection of a fraction of the total energy 
[$\sim 7 \times 10^{50}  
(f_{\rm rad}/0.1)^{-1}$ erg: see above] into the nebula within 
$\sim t_0$ would naturally explain large $v_0$. 

\item[(4)]
A connection to other SNe Ib/c is interesting. 
A typical pulsar with $B_{\rm mag} \sim 10^{12}$ gauss and $P_0 \sim 30$ ms 
produces only negligible contribution to the light curve compared to 
the $^{56}$Ni/Co energy input during the first few years.  
In the pulsar energy input scenario, 
large $L_0$ and small $t_0$ (thus large $B_{\rm mag}$ and small $P_0$) 
are required to make the light curve doubly peaked 
as seen in SN 2005bf. Such that, the model is consistent with singly 
peaked light curves of other SNe Ib/c, which are believed to leave 
a typical neutron star (except probably the Gamma-Ray Burst related SNe). 

Another interesting implication is related to SN 2006aj associated with 
an X-Ray Flash (XRF) 060218 (Pian et al. 2006). 
SN2006aj/XRF060218 is suggested to be driven by a neutron star formation, 
presumably a magnetar, through observations at early phases 
(Mazzali et al. 2006) and at late phases (Maeda et al. 2007). 
SN 2006aj showed a singly-peaked light curve as is similar to 
other SNe Ib/c explained by usual $^{56}$Ni heating scenario. 
This behavior is explained, if the newly born neutron star in SN 2006aj 
has even larger $B_{\rm mag}$ and/or smaller $P_0$ than in SN 2005bf. 
In this case, the characteristic time scale ($t_0$) 
of the high energy input becomes as small as $t_0 \lsim 10^{\rm d}$ 
since the dipole radiation is scaled as $L_{0} \propto 
B_{\rm mag}^{2} P_0^{-4}$. 
A magnetar may also spin down much shorter than $t \sim 1^{\rm d}$ without emitting 
electromagnetically, if it blows a massive wind (Thompson, Chang, \& Quataert 2004). 
Most of the emission is then consumed 
by adiabatic lose because of high density in such an early epoch. 
Such that, the contribution of the pulsar energy input 
to the light curve is negligible in this case, and a part of 
the pulsar energy input is transferred to the SN ejecta. 
This may explain $E_{51} \sim 2$ in SN 2006aj. 

Thus, we suggest that SN 2005bf is an event linking 
usual SNe Ib/c and SN 2006aj/XRF 060218. In our proposed scenario, 
these are connected by the formation of a neutron star with different 
$B_{\rm mag}$ and $P_0$.

\end{itemize}

Although the choice of the model parameters look reasonable, 
there is a caveat. 
Since the nature of young pulsar is still in active debate, 
more detailed study of the magnetar hypothesis is necessary.  
For example, even the very basic assumption in this model, i.e., 
whether the pulsar wind nebula is formed within a few days 
since the explosion, is still under debate (e.g., Fryer, Colgate, \& Pinto 1999). 

\section{CONCLUSIONS}

\subsection{Blob Model for the First Peak and the Nebular Epoch} 

In this paper, we have presented the results from spectroscopic 
and photometric observations of SN 2005bf at $t \sim 300^{\rm d}$. 
Our theoretical considerations are summarized as follows. 

\begin{itemize}
\item[(1)] 
The faint nebular emission, composed of blueshifted emission lines 
([OI], [CaII], [FeII]), can be understood if 
the emission in the late phases is dominated by a  
low mass blob with $M_{\rm ej, neb} \sim 0.1 - 0.4\Msun$, 
$M$($^{56}$Ni)$_{\rm neb}$ $\sim 0.02 - 0.06\Msun$.
\item[(2)] 
The blueshift is reproduced either by a unipolar blob 
with the center-of-mass velocity $v \gsim 2,000 - 5,000$ km s$^{-1}$, or  
by self-absorption of the optical light as seen in SNe 1990I and 1987A. 
\item[(3)]
The emission line profiles 
from different elements suggest that 
the blob in itself has layered structure. 
\item[(4)]
The optical luminosities at the first peak ($t \sim 20^{\rm d}$) 
and at the nebular phases ($t \sim 300^{\rm d}$) 
are consistent with the emission from this blob. 
\item[(5)]
The line ratios suggest the abundance pattern similar to what is expected 
from progenitor stars with $M_{\rm ms} \sim 20 - 25\Msun$, 
as is consistent with the lower end of the estimate 
given by the previous works (Tominaga et al. 2005). 
\end{itemize} 

It should be mentioned that recently a new paradigm  
is entering into the scene of core-collapse physics, 
either standing accretion shock instability or g-mode oscillation of the 
newly born neutron star. Some models do predict 
unipolar supernova explosions (Burrows et al. 2006). 
SN 2005bf could be the first extreme example of this kind of explosions. 

\subsection{Energy Source for the Second Peak}

At the nebular phases, SN 2005bf turns out to be extremely 
faint ($R = 24.4$ mag at $t \sim 273^{\rm d}$). 
It is dominated by the contribution from the low-mass blob. 
The bulk of the emission expected from 
$M$($^{56}$Ni)$_{\rm peak}$ $\sim 0.3\Msun$, as is 
derived from the second peak luminosity at $t \sim 40^{\rm d}$, is missing. 
Where is the missing emission is a question we tried to answer in 
this paper. 

As the energy source should be buried deep in the ejecta, 
the peak date is determined by the diffusion time scale. 
Therefore, the ejecta mass and the energy derived 
by the previous works should give a good estimate even if the 
energy source responsible to the peak luminosity is different. 
In conclusion, the main sequence mass is $M_{\rm ms} \sim 20 - 25\Msun$, 

Four possibilities are examined in \S 5: 
(1) accelerated gamma-ray and positron escape, 
(2) almost $100 \%$ shielding of the optical light, 
(3) fallback of $^{56}$Ni-rich materials, and 
(4) possible central object's activity. 
Among these, the first three possibilities have difficulties. 

(1) No physically reasonable mechanism is found to reduce the gamma-ray and 
positron opacity (\S 5.1). (2) The extreme self-absorption scenario 
looks to be inconsistent with the detail of the observations (\S 5.2). 
(3) The fallback scenario is found to be difficult to work, unless 
some additional mechanisms (e.g., delayed fallback by disk accretion) 
could rescue the situation (\S 5.3). 

One can still consider a combination of some of them. For example, 
assume that gamma-rays but not positrons can escape effectively 
and that the optical output is reduced by a factor of 10 -- 15 by self--absorption. 
This is more extreme than in Model B and other dust-forming SNe, 
but less than examined in \S 5.2. 
Then the late-time luminosity could be reproduced (see the light 
curve with small $\kappa_{\gamma}$ in Figure 8). A question 
here is if these phenomena, each of which is unusual, 
can by chance take place together. 

\subsection{Magnetar Hypothesis}
The last possibility, the central remnant activity, yields a reasonable 
fit to the light curve if the central remnant is a magnetar with 
$B_{\rm mag} \sim 10^{14 - 15}$ gauss and $P_0 \sim 10$ ms. 
The scenario has advantages compared to other models as summarized in the following. 

\begin{itemize} 

\item[(1)] 
The magnetar hypothesis can explain two peculiar features 
in the light curve in the same context. 
The rapid declining at $t \sim 60^{\rm d}$ and the 
faintness at $t \sim 300^{\rm d}$ are essentially difficult 
in the standard $^{56}$Ni/Co heating scenario. 
These two could be explained 
by combination of two (very) peculiar natures, such as the reduced $\gamma$-ray 
opacity for the former and the huge dust extinction for the latter, but 
in the magnetar hypothesis these are attributed to the single physical 
reason. 

\item[(2)] 
The blueshift of the nebular emission lines could be related to the pulsar kick. 
The blueshift can be interpreted as ejection of the unipolar blob with $v \lsim 
2,000$ km s$^{-1}$ (Model A). With $M_{\rm neb} \sim 0.1\Msun$ in the blob, 
the newly formed neutron star with $\sim 1.4\Msun$ would have the kick velocity of 
$v_{\rm kick} \gsim 140$ km s$^{-1}$. 

\item[(3)]
The scenario is compatible to relatively large $E_{51} \sim 1 - 1.5$, 
as the magnetar activity could also increase the ejecta kinetic energy. 
It could also be compatible to the estimated mass, $M_{\rm ms} \sim 20 - 25\Msun$, 
which is close to the upper limit for the neutron star formation. 

\item[(4)] 
The rarity of SN 2005bf-like supernova is consistent with 
the rarity of a magnetar, although there are observational biases in 
the search of neutron stars. 
\end{itemize} 

Summarizing, we suggest that SN 2005bf is driven by 
a strongly magnetized neutron star (a magnetar), 
being the birth place of a soft gamma-ray repeater or an anomalous 
X-Ray pulsar. 
In our scenario, SN 2005bf is an event which links usual SNe Ib/c and 
SN 2006aj/XRF 060218: As the magnetic activity and/or the spin frequency increases, 
the resulting SN becomes usual SNe Ib/c (a typical neutron star, whose contribution 
to the light curve is negligible), SN 2005bf-like supernova (for which 
the magnetar makes the doubly-peaked light curve), and finally SN 2006aj/XRF 060218-like 
high energy transient (again the light curve becomes singly peaked, since 
the magnetar activity is consumed to produce the high energy transient and 
to increase the ejecta kinetic energy).

\acknowledgements
The authors would like to thank Brian Schmidt, 
Philipp Podsiadlowski, and Sergei Blinnikov for 
useful discussion. The authors also thank 
Jinsong Deng, Elena Pian, and Abouazza Elmhamdi 
for constructive comments. 
The authors also thank all the staff at the Subaru observatory 
for their excellent support of the observations. 
This research has been supported in part by the 
National Science Foundation under Grant 
No. PHY99-07949, and by the Grant-in-Aid 
for Scientific Research 
(17030005, 17033002, 18104003, 18540231 for K.N.) and the 
21st Century COE Program (QUEST) from the JSPS and MEXT of Japan.
K.M. is supported through the JSPS 
(Japan Society for the Promotion of Science) Postdoctoral Fellowships 
for Research Abroad. N.T. is a JSPS Research Fellow. 

\appendix

\section{Light Curves of Supernovae with Fallback}

The light curves of supernovae undergoing fallback 
(Fig. 10) are computed as follows. 
They are computed using the same code described in \S 4. 
The same ejecta model with $M_{\rm ej, peak} = 7.3\Msun$ and 
$E_{{\rm peak}, 51} = 1.3$ is adopted, except for 
the $^{56}$Ni distribution. 
Here we set the $^{56}$Ni mass fraction above $1,600$ km s$^{-1}$ 
zero since we are not concerned with the first peak (see \S 5.3). 

We remove the ejecta materials (including $^{56}$Ni)
from the innermost region as a function of time.  
$M_{\rm ej}$ is decreased according to 
\begin{equation} 
\dot M_{\rm ej} (t) = \dot M_{\rm ej} (t_{\rm acc}) 
\times \left(\frac{t}{t_{\rm acc}}\right)^{-\frac{5}{3}} \ , 
\end{equation} 
where $t_{\rm acc}$ is the date when the fallback is assumed to begin. 
Thus, $M_{\rm ej} (t) = M_{\rm ej, peak}$ for $t \le t_{\rm acc}$. 
This temporal dependence is expected in the limit of 
negligible pressure support and confirmed by numerical 
calculations (Woosley \& Weaver 1995; also see Appendix B). 
The model parameters are as follows: 
\begin{itemize} 
\item $t_{\rm acc} = 40^{\rm d}$ and $\dot M_{\rm ej} (t_{\rm acc}) = 1.2 \times 10^{-2} \Msun$ 
d$^{-1}$, and 
\item $t_{\rm acc} = 20^{\rm d}$ and $\dot M_{\rm ej} (t_{\rm acc}) = 2.0 \times 10^{-2} \Msun$ 
d$^{-1}$. 
\end{itemize} 
The corresponding histories of the mass accretion are 
shown in Figure 11. 

\section{Spherical Fallback}

Figure 11 for the histories of the fallback mass accretion rate 
is computed as follows. 
A set of 1D explosion simulations are performed 
for a $7.3\Msun$ He core of a star with $M_{\rm ms} = 25\Msun$, 
using a 1D PPM (piecewise parabolic method) hydrodynamic code. 
We have varied the explosion energy which is injected at $M_r = 1.8\Msun$ 
and investigated the relation among the final kinetic energy ($E$), 
the amount of the fallback materials ($M_{\rm acc}$), and the timescale of the fallback 
($t_{\rm acc}$). 
Figure 11 shows the histories of the fallback obtained in the simulations. 
Also shown in Figure 11 are the histories of the fallback assumed to 
compute the light curves in Figure 10. 
By comparing the shapes of the curves in Figure 11, 
it is seen that the fallback temporal dependence used in 
the light curve computations ($\dot M_{\rm ej} \propto 
t^{-5/3}$) is a good approximation for the spherical hydrodynamic fallback. 

From these simulations, we find the following relation.  
Smaller $E$ results in larger $M_{\rm acc}$ and smaller $t_{\rm acc}$.  
The light curve fitting requires $t_{\rm acc}$ much larger than 
that obtained by the hydrodynamic 
simulations. Also, $E_{{\rm peak}, 51} = 1 - 1.5$ (Table 1) 
is too large to make the fallback effectively work. 

We set the position of the energy injection at $M_r = 1.8\Msun$ in this 
examination. If we take larger $M_r$ for the injection, 
then the binding energy in the surrounding layers is smaller. 
Then it acts like a less massive star 
as long as the fallback is concerned. 
A less massive star experiences 
the fallback for smaller $E$ (see Iwamoto et al. 2005). 
According to our simulations, the final kinetic energy 
dividing the fallback and no-fallback is $E_{51} \sim 0.2$ for $M_r = 3.6\Msun$, 
which is smaller than $E_{51} \sim 0.7$ for $M_r = 1.8\Msun$. 
For $M_r = 3.6\Msun$, $t_{\rm acc}$ can be a bit longer than for $M_r = 1.8\Msun$, 
but still $t_{\rm acc} \lsim 10^{\rm d}$. 
Because $E_{{\rm peak}, 51} \sim 1 - 1.5$, 
the larger $M_r$ makes the fallback scenario less 
likely to be realized. In sum, changing $M_r$ does not solve the problem.

\end{document}